\documentclass[fleqn,usenatbib]{mnras}

\usepackage{newtxtext,newtxmath}

\usepackage[T1]{fontenc}
\usepackage{subcaption}
\usepackage{booktabs}
\usepackage{tabularx}

\usepackage{graphicx}	
\usepackage{amsmath}	
\usepackage{float}
\usepackage{cuted}
\usepackage{xcolor}

\title[Degeneracy between satellites and baryonic feedback]{Interpreting the stacked kinetic SZ signal II: the degeneracy between satellite galaxies and baryonic feedback}
\author[Ondaro-Mallea et al.]{
Lurdes Ondaro-Mallea$^{1,2}$,\thanks{E-mail: lurdes.ondaro@ast.cam.ac.uk}
Boryana Hadzhiyska$^{1,2}$
and Raul E. Angulo$^{3,4}$
\\
$^{1}$Institute of Astronomy, University of Cambridge, Madingley Road, Cambridge CB3 0HA, United Kingdom\\
$^{2}$Kavli Institute for Cosmology Cambridge, Madingley Road, Cambridge CB3 0HA, UK\\
$^{3}$ {Donostia International Physics Center, Manuel Lardizabal Ibilbidea, 4, 20018 Donostia, Gipuzkoa, Spain}\\
$^{4}$IKERBASQUE, Basque Foundation for Science, 48013, Bilbao, Spain \\
}
\date{Accepted XXX. Received YYY; in original form ZZZ}

\pubyear{2015}

\begin{document}
\label{firstpage}
\pagerange{\pageref{firstpage}--\pageref{lastpage}}
\maketitle
\begin{abstract}
The kinetic Sunyaev-Zel'dovich (kSZ) signal stacked around a galaxy sample traces the gas around halos and is a powerful probe of baryonic feedback. Its interpretation, however, depends on the satellite galaxies in the sample. Using the FLAMINGO hydrodynamical simulations and DESI-like samples of luminous red galaxies (LRGs),  bright galaxies survey (BGS), and emission-line galaxies (ELGs), we show that satellites and feedback have similar effects: both suppress the small-scale signal. This degeneracy arises because stacking on satellites, which are offset from the centre of their host haloes, mimics the extended gas distribution produced by feedback. Although satellites make up only 20-30\% of each sample, they account for 50-80\% of the signal because they occupy more massive halos than centrals; thus their imprint on the stacked profiles can be as large as that of feedback. At the precision of current DESI$\times$ACT data, neglecting satellites in the modelling biases the inferred gas fractions low, by up to a factor of two, and overestimates the feedback strength, particularly for the LRG and BGS samples. Remaining agnostic about the satellite population avoids this bias but reduces the constraining power, so robust constraints would require knowing the satellite fraction to better than 10\%, for instance from lensing and clustering data of the same galaxy samples. Finally, a further contribution from satellites will become statistically significant with upcoming surveys: satellites trace the anisotropic gas distribution around their host haloes, an effect that contributes 10\% of the signal, and will likely require simulation-based models to capture. 
\end{abstract}


\section{Introduction}
The uncertain mapping of baryons and dark matter on small scales is one of the largest limiting factors for extracting robust cosmological information from Stage IV surveys, particularly weak gravitational lensing. While predicting the baryon distribution from first-principles is notoriously hard, direct observations of the diffuse gas around haloes offer a very valuable insight into the problem. 

The diffuse gas can be probed through its X-ray emission and, at millimetre wavelengths, through the thermal and kinetic Sunyaev-Zel'dovich (SZ) effects, which arise from the scattering of CMB photons by free electrons in ionised gas around clusters and galaxies \citep[e.g.][]{Sunyaev:1972,Sunyaev:1980,Mroczkowski:2019}. X-ray and thermal SZ (tSZ) measurements have long provided significant constraints on the gas in massive galaxy clusters \citep[e.g.][]{Vikhlinin:2006,Hoekstra:2015,Eckert:2016,Lovisari:2020}. In contrast, the kinetic SZ (kSZ) effect is substantially weaker, and statistically significant measurements have only recently become possible through velocity-weighted stacking on large galaxy samples \citep{Li:2014}. Combining large galaxy surveys such as BOSS \citep{Dawson2013BOSS} and DESI \citep{DESI:2016} with high-resolution CMB observations has now enabled highly significant kSZ detections \citep{Schaan:2016,Schaan:2021,Hadzhiyska:2021,RiedGuachalla:2025,Qu:2026,Hadzhiyska:2026}. These measurements extend the observed regime to galaxy group scales and higher redshifts than traditional X-ray and tSZ cluster measurements \citep[see also][for tSZ stacks around groups]{Liu:2025,Siegel:2026b}. 

Combining these probes has revealed an apparent discrepancy. The first interpretations of the kSZ measurements \citep[e.g.][]{Hadzhiyska:2024, Bigwood:2024, McCarthy:2025b} infer significantly lower gas fractions than those obtained from analyses of X-ray observations of galaxy clusters, which favoured relatively mild baryonic feedback \citep[e.g.][]{Grandis:2024} that is reproduced by many hydrodynamical simulations \citep[e.g.][]{Schaye:2015, McCarthy:2017, Pillepich:2018, Dave:2019, Parkmor:2023, Schaye:2023}. Recent \textit{eROSITA} X-ray measurements of optically selected galaxy groups have likewise reported low gas fractions \citep{Popesso:2026}. However, the picture is not homogeneous: other kSZ \citep{Liu:2026} and X-ray analyses \citep{Seppi:2026, Khalil:2026, McDonald:2026} find gas fractions consistent with a milder feedback scenario.

The origin of these discrepancies is actively debated. They could arise from extrapolations between different halo mass and redshift regimes, since traditional X-ray measurements typically target more massive, lower-redshift haloes than those probed by kSZ stacks of DESI galaxy samples \citep[e.g.][]{Lucie-Smith:2025,Sharma:2026}, even though kSZ stacks are progressively covering a broader range of masses and redshifts, already partially overlapping with those probed by X-ray measurements \citep{McCarthy:2025b}. Selection effects may also play a role, particularly for X-ray-selected samples which may preferentially select the gas-rich haloes at a given halo mass \citep[e.g.][]{Sun:2009,Lovisari:2015,Bulbul:2024}. Alternatively, the overall amplitude of kSZ measurements is potentially uncertain, as it scales with the cross-correlation coefficient between true and reconstructed velocities that has to calibrated through simulations \citep{RiedGuachalla:2023,Hadzhiyska:2023}. Finally, the discrepancy could result from modelling assumptions used to interpret the measurements. In a previous companion paper, we investigated the role of non-linearities on the small-scale gas velocity field and its coupling to the reconstructed velocities, finding that they change the shape of the small-scale profile at $\approx 1\sigma$ for current DESI DR2$\times$ACT measurements, and will become statistically significant in the near future \citep{OndaroMallea:2026}.

In this paper, we investigate one such potential modelling systematic: the presence of satellite galaxies in the samples used for kSZ stacking. Stacking on satellite galaxies is known to suppress the small-scale kSZ signal \citep[e.g.][]{Hadzhiyska:2023b,McCarthy:2025b,Siegel:2025,Bigwood:2025}, and different analyses have adopted different approaches to account for this effect. Analyses based on analytical models often assume that the galaxy samples consist entirely of centrals \citep{Amodeo:2021,Bigwood:2024,Kovac:2025,Siegel:2026}\footnote{\cite{Siegel:2026} tests the validity of central-only modelling and imposes scale cuts to remain within the validated regime. We discuss this approach in Section \ref{sec:fixed_n}.}, or assume a simple HOD based analytical model \citep{Hadzhiyska:2026,Qu:2026}. In contrast, analyses based on hydrodynamic simulations typically select galaxies above a given stellar-mass threshold (including centrals and satellites), chosen to reproduce either the mean number density \citep{Hadzhiyska:2024,RiedGuachalla:2025} or mean halo mass of the sample measured with CMB lensing \citep{Hadzhiyska:2025}, or the galaxy-galaxy lensing measurement of the same sample \citep{McCarthy:2025b,Siegel:2025,Bigwood:2025}.

To quantify the role of satellite galaxies, we will use the FLAMINGO simulations \citep{Schaye:2023,Kugel:2023,Helly:2026} and construct mock kSZ measurements for realistic DESI-like galaxy selections. We will then systematically investigate the interplay between satellite galaxies in the sample and the surrounding gas distribution, and quantify the extent to which neglecting or mis-modelling satellites can bias the inferred gas fractions and constraints on baryonic feedback. Our main findings are: 1) neglecting satellites in the modelling biases the inferred halo gas fractions at the precision of current DESI $\times$ ACT data, with satellite miscentering being the dominant effect; and 2) with upcoming data, higher-order contributions from the anisotropic distribution of gas and satellites will become significant, likely requiring simulation-based models to capture.

The paper is organised as follows. In Section \ref{sec:background}, we review the stacked kSZ estimator and introduce our analysis framework. Section \ref{sec:methods} describes the FLAMINGO simulations, galaxy selections, and mock kSZ measurements. In Section \ref{sec:sat_geom}, we study the optical depth around satellite galaxies and its qualitative differences from that around centrals. In Section \ref{sec:cen_sat_bar}, we study how baryonic effects change the optical depth profile around central and satellite galaxies separately. Section \ref{sec:sat_bar_deg} quantifies the degeneracy between these effects and shows that neglecting satellites in the theoretical modelling can bias constraints on baryonic feedback. In Section \ref{sec:implications}, we discuss several approaches to account for the satellite population and modelling requirements. Section \ref{sec:discussion} discusses the broader implications of our results in the context of previous work. Finally, Section \ref{sec:conclusions} summarises our main findings and concludes.

\section{Background}\label{sec:background}
In this section we write down the estimator for the stacked kSZ signal around a galaxy sample, and schematically summarise what it is sensitive to as a function of scale. For more details, we refer to a companion paper \cite{OndaroMallea:2026}.

The kSZ effect is a secondary anisotropy of the cosmic microwave background (CMB), produced by the Thomson scattering of CMB photons off free electrons with a non-zero line-of-sight peculiar velocity relative to the CMB rest frame \citep{Sunyaev:1972,Sunyaev:1980}. The resulting temperature fluctuation is proportional to the line-of-sight integral of the free-electron momentum field. 
To extract this signal, we stack the CMB temperature map, $T_{\rm CMB}$, around a galaxy sample, weighted by an estimate of the line-of-sight velocity of the gas around the galaxy, $v_{\rm stack}$, 
\begin{equation}\label{eq:ksz_estimator}
    T_{\rm ksz}(r_{\rm p}) = \left \langle T_{\rm CMB}(r_{\rm p}) \, v_{\rm stack} \right \rangle,
\end{equation}
where $r_{\rm p}$ is the projected distance from the galaxies and the average goes over all galaxies in the sample.

To leading order, the stacked signal measures the mean optical depth profile around the galaxy sample,
\begin{equation}
T_{\rm ksz}(r_{\rm p}) \approx \langle \tau(r_{\rm p}) \rangle v_{\rm rms}^{h} r,
\end{equation}
where $\langle \tau(r_{\rm p}) \rangle$ is the mean optical depth, $v_{\rm rms}^{h}$ is the rms line-of-sight velocity of the host haloes, and $r$ is the cross-correlation coefficient between gas velocity and stacking velocity. We assume $v_{\rm rms}^{h}$ and $r$ are known and focus on the mean optical depth profile. Higher-order scale-dependent corrections are typically $\approx 10\%$ and are neglected here \citep{OndaroMallea:2026}. 

After stacking, we apply a compensated aperture photometry (CAP) filter to suppress CMB primary fluctuations:
\begin{equation}\label{eq:tau_cap}
\tau^{\rm CAP}(r_{\rm p}) = \tau(<r_{\rm p}) - \left[ \tau(<\sqrt{2}r_{\rm p}) -\tau(<r_{\rm p}) \right].
\end{equation}
The CAP filter measures the \textit{excess} projected signal within the aperture relative to the adjacent annulus. Unless otherwise stated, all optical depth profiles below are CAP-filtered, and we omit the superscript ``CAP''.

The electron number density is given by the ionised gas density up to a normalising factor. Moreover, in the low-redshift Universe the gas is almost fully ionised, except in the cold, dense regions of galaxies where neutral gas can survive. Thus, the stacked kSZ estimator roughly traces the mean gas distribution around the galaxy sample.

The physical interpretation of such profiles is clearest when separating large and small scales. We summarise the main physical picture in Table~\ref{tab:physical_summary}.

\subsection{Large scales}

On scales much larger than the virial radius, the profile approaches a plateau as the aperture encloses nearly all of the gas associated with the halo. Its amplitude is therefore primarily sensitive to the total gas mass, scaling approximately with halo mass and the universal baryon fraction.

Two second-order effects modify this simple picture. First, a fraction of the baryons is locked into stars \citep{Giodini:2009,Zhang:2011,Reyes:2012,Gonzalez:2013,Kravtsov:2018,Chiu:2018,Akino:2022,Grandis:2024}, thus the optical depth profile does not plateaus at the universal baryon fraction, but at the universal baryon fraction minus the stellar mass fraction of the mean halo. Second, correlated large-scale structure (the two-halo term) contributes gas beyond the virial radius, producing a gradual increase in the profile. 

On large angular scales, $\theta \gtrapprox  3-4 \, \rm arcmin$, the S/N is limited because primary CMB anisotropies dominate. 

\subsection{Small scales}
On scales comparable to and smaller than the virial radius, the optical depth profile probes the distribution of gas relative to the galaxies used as stacking centres.

More massive haloes produce a higher-amplitude and more extended profile, reaching the plateau at larger radii. At fixed stellar mass, this also occurs when satellite galaxies are included, since satellites typically reside in more massive haloes than central galaxies (see Table~\ref{tab:sample_properties}).

At fixed halo mass, and therefore fixed large-scale amplitude, the profile shape is primarily affected by baryonic effects and the choice of stacking centres.

First, some baryonic mass, initially in the form of gas, is locked into stars, reducing the gas density on small scales. Baryonic feedback then redistributes some of the remaining gas from the inner halo to larger radii, suppressing and broadening the gas profile. Second, stacking on satellite galaxies similarly suppresses and broadens the profile, because the signal is measured around satellite positions rather than at the halo centre, where the gas density peaks.

Most of the constraining power of the stacked kSZ measurements comes from small angular scales of $\theta \lessapprox 3 \, \rm arcmin$, which correspond to the 1-halo or 1 to 2-halo regime, depending on the halo mass and redshift of the galaxy sample. 

\subsection{Analysis strategy}

The primary goal of stacked kSZ analyses is to constrain the relation between gas and dark matter distributions as a function of halo mass and redshift. This is commonly summarised by the halo gas fraction,
\begin{equation}
 f_{\rm gas}(M)=\frac{M_{\rm gas}}{M},
\end{equation}
 where $M_{\rm gas}$ is the gas mass within halo radius $R$, and $M$ is the total mass within the same radius. We emphasise that this is a summary statistic: kSZ measurements carry information of the whole gas profile as a function of radius. 
 
The gas fraction is expected to reach a minimum in group-scale haloes, where AGN feedback is most efficient ($M \approx 10^{13}\,h^{-1}M_\odot$), and to increase towards higher halo masses, approaching the universal baryon fraction in massive clusters where feedback can no longer efficiently expel gas \citep[e.g.][]{Ettori:2006,Haider:2016,Springel:2018,Zinger:2020,Lucie-Smith:2025,Pontzen:2026,Peiris:2026}. 
A robust physical interpretation therefore requires an estimate of the characteristic halo mass of the galaxy sample, since the inferred gas fraction is meaningful only when interpreted at a given halo mass.

In this work, we fix the halo mass of each galaxy sample to the values measured observationally from external datasets (see Section \ref{sec:galaxy_samples} and Table \ref{tab:sample_properties} for a summary of the sample properties). We then focus on two of the remaining processes affecting the small-scale signal, namely baryonic feedback and satellite galaxies in the sample, and the degeneracy between them.

\begin{table}
\centering
\caption{Physical interpretation of the stacked optical depth profile on large and small scales. In this work, we fix the halo mass of the sample, thereby fixing the large-scale amplitude, and focus on the remaining degeneracies that shape the small-scale profile, namely baryonic feedback and satellite galaxies.}
\begin{tabularx}{\columnwidth}{@{}p{0.6cm} X X c@{}}
\toprule
\textbf{Scale} & \textbf{What is measured} & \textbf{Physical dependencies} & \textbf{S/N} \\
\midrule
Large
& Amplitude (plateau) of the stacked optical depth profile
& Halo mass (dominant); two-halo term, stellar fraction (secondary)
& Low \\
\addlinespace
Small
& Shape of the stacked optical depth profile
& Baryonic feedback, satellite miscentering, halo mass / size, stellar fraction
& High \\
\bottomrule
\end{tabularx}
\label{tab:physical_summary}
\end{table}

\section{Methods}\label{sec:methods}

In this section we briefly introduce the simulations, halo and subhalo finding, galaxy selections, and the estimator we use. For more details we refer to the companion paper \cite{OndaroMallea:2026}.

\subsection{Simulations}\label{sec:flamingo_sims}

We use the FLAMINGO suite of cosmological hydrodynamical simulations \citep{Schaye:2023,Kugel:2023,Helly:2026}, at the fiducial resolution run. The suite provides a fiducial model, denoted as $f_{\rm gas}$, calibrated to the pre-\textit{eROSITA} X-ray gas fractions of clusters and to the low-redshift stellar mass function, alongside a set of runs that deviate systematically from the observed relations, either in cluster gas fractions or the stellar mass function \citep[see][for the calibration data and strategy]{Kugel:2023}.  In this paper, we focus on the deviations from the fiducial cluster gas fractions, denoted as $f_{\rm gas} \pm N\sigma$.

\subsection{Halo and subhalo finding}\label{sec:halo_subhalo_finding}

The halo catalogues are constructed by running a 3D friends-of-friends (\texttt{FoF}) algorithm \citep[e.g.][]{Press:1982} on the full particle field. Substructures are then identified with the \texttt{HBT-HERONS} finder, which uses the assembly history of each substructure \citep{Forouhar:2025}. The position of each substructure is taken to be that of its most bound particle, i.e.\ an approximation to the minimum of the potential.

Substructure finding, and the classification of a given substructure as a central or a satellite, can depend strongly on the design and choices of the adopted algorithm \citep{Forouhar:2025}. In Appendix \ref{sec:satellite_definition} we illustrate this issue and test how our results change for other subhalo finders and different satellite definitions. The exact numbers change slightly, but the overall picture is consistent across different choices.

We use the halo properties computed with Spherical Overdensity and Aperture Processor \citep{McGibbon:2025}. Throughout, we define the halo radius $R_{\rm 200m}$ as the radius enclosing a mean density 200 times the mean density of the Universe, and the host halo mass $M_{\rm 200m}$ as the mass enclosed within it.

\subsection{Galaxy samples}\label{sec:galaxy_samples}
\begin{figure}
    \centering
    \includegraphics[width=0.8\linewidth]{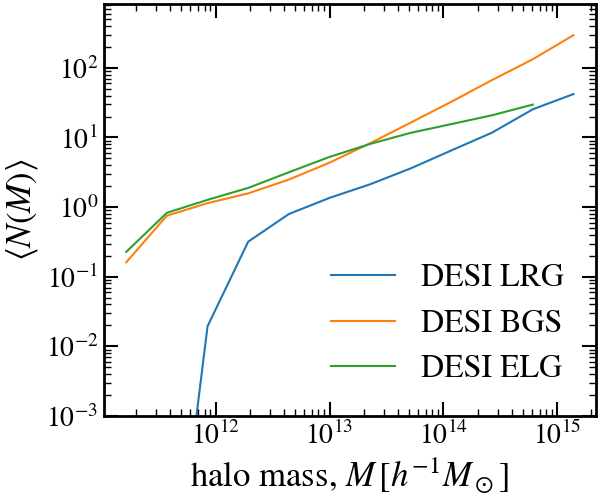}
    \caption{Halo occupation distributions of the three DESI-like mock samples: LRGs (blue), BGS (orange), and ELGs (green). The stellar-mass-selected samples (LRGs and BGS) exhibit pronounced high-halo-mass tails, whereas the ELG sample has a broader halo mass distribution. The high-mass end of each distribution is dominated by satellite galaxies.}
    \label{fig:hod}
\end{figure}

We construct three DESI-like galaxy samples, chosen to mimic the populations used in the highest-significance stacked kSZ measurements to date: luminous red galaxies (LRGs) at $z=0.7$ \citep{Zhou:2023}, the bright galaxy sample (BGS) at $z=0.25$ \citep{Hahn:2023}, and emission line galaxies (ELGs) at $z=1.1$ \citep{Raichoor:2023}. We require the samples to have the mean halo mass inferred from CMB lensing measurements for the LRG and BGS samples \citep{Hadzhiyska:2026,Qu:2026}, or implied by HOD models of ELG clustering \citep{Garcia-Quintero:2025}.

\begin{table*}
\centering
\caption{Halo occupation properties of the galaxy samples. Halo masses are given as $\log_{10}(M_{200\mathrm{m}}/h^{-1}M_\odot)$. The mean miscentering, $\langle d\rangle/\langle R_{200\mathrm{m}}\rangle$, is the mean satellite distance from the halo centre normalised by the mean halo radius, and $f_{\rm sat}$ is the satellite fraction.}
\begin{tabularx}{\textwidth}{@{}l *{8}{>{\centering\arraybackslash}X}@{}}
\toprule
\textbf{Sample} &
\textbf{$z$} &
\textbf{$\log\langle M\rangle$} &
$\log\langle M_\star \rangle$&
\textbf{$\log\langle M_{\rm cen}\rangle$} &
\textbf{$\log\langle M_{\rm sat}\rangle$} &
\textbf{$\langle M_{\rm sat}\rangle/\langle M_{\rm cen}\rangle$} &
\textbf{$\langle d\rangle/\langle R_{\rm 200m}\rangle$} &
\textbf{$f_{\rm sat}$} \\
\midrule
DESI LRG & 0.70 & 13.20 & 11.13 & 12.84 & 13.64 & 6.32 & 0.79 & 0.24 \\
\addlinespace
DESI BGS & 0.25 & 13.24 & 10.57 &12.20 & 13.70 & 31.66 & 0.67 & 0.32 \\
\addlinespace
DESI ELG & 1.10 & 12.41 & 10.33 & 11.93 & 12.81 & 7.65 & 0.77 & 0.30 \\
\bottomrule
\end{tabularx}
\label{tab:sample_properties}
\end{table*}

The main properties of the resulting samples are summarised in Table~\ref{tab:sample_properties}. The BGS and LRG samples have very similar mean halo masses of $\log_{10}M \simeq 13.2$. However, the galaxies in the BGS sample have lower stellar masses, reproducing the observational trends reported by \citet{Hadzhiyska:2026}. The BGS sample is the lowest redshift sample and has the highest satellite fraction, as expected from hierarchical structure formation.

Figure~\ref{fig:hod} shows the halo occupation distributions (HODs) of the three samples. The LRG and BGS samples have pronounced high-mass halo tails mainly populated by satellite galaxies. Instead, the distribution of ELGs is flatter on halo mass.

\subsection{Estimator}\label{sec:estimator}

We compute the CAP-filtered mean optical depth around the galaxy samples by measuring the projected gas mass around each galaxy using a modified version of the publicly available \texttt{halotools} package\footnote{\url{https://github.com/astropy/halotools}} \citep{Hearin:2017}, and later applying the CAP filter (Eq. \ref{eq:tau_cap}). 

Once the CAP filter is applied, the profiles converge after integrating over $\approx 10\,h^{-1}\,\mathrm{Mpc}$ along the line of sight \citep[see][]{OndaroMallea:2026}. This is well below the velocity coherence length of $\approx 50\,h^{-1}\,\mathrm{Mpc}$, so no additional line-of-sight cut is required.

In Sections \ref{sec:sat_geom} and \ref{sec:cen_sat_bar}, we present the profiles as a function of comoving projected distance from the galaxy positions, without applying an instrumental beam. In Section~\ref{sec:sat_bar_deg}, we additionally express the profiles in angular units and convolve them with a Gaussian beam of $\mathrm{FWHM}=1.6\,\mathrm{arcmin}$, corresponding to the ACT beam.

\section{Optical depth around satellite galaxies}\label{sec:sat_geom}
In this section, we study the optical depth around satellite galaxies, and its qualitative differences from that around central galaxies.

The optical depth around satellite galaxies is given by the convolution of the host halo optical depth with the projected satellite position within the host. We derive this expression, examine its components, and quantify their physical origins and dependence on the satellite radial distribution. To do so, we analyse the optical depth around the satellite galaxies in the DESI LRG-like sample.

\subsection{Decomposition of the satellite optical depth profile: isotropic and anisotropic components}\label{sec:sat_dec}

Let the host optical depth $\tau_{\rm host}(\vec{r}_{\rm p})$ be a function of projected
position from the host centre $\vec{r}_{\rm p}$, which for a non-spherical halo depends
on the full vector and not only on its modulus,
\begin{equation}
\vec{r}_{\rm p} = r_{\rm p}\,\hat{n}(\theta),
\end{equation}
with $\hat{n}(\theta) \equiv (\cos\theta, \sin\theta)$, and $\theta$ is the azimuthal angle relative to the host halo centre. Its azimuthal average reads
\begin{equation}
\bar{\tau}_{\rm host}(r_{\rm p}) \equiv \frac{1}{2\pi}\int_{0}^{2\pi}\!{\rm d}\theta\;
\tau_{\rm host}\!\left(r_{\rm p}\,\hat{n}(\theta)\right).
\end{equation}
The projected position of a satellite with respect to the host halo centre is
\begin{equation}
\vec{r}_{\rm p}^{\,\rm sat} = r_{\rm p}^{\rm sat}\,\hat{n}(\theta^{\rm sat}).
\end{equation}
Recentring on the satellite and averaging over the azimuthal angle $\phi$ relative to the satellite position, the optical depth profile around a satellite reads
\begin{equation}\label{eq:sat_ksz}
\tau_{\rm sat}(r_{\rm p})
= \frac{1}{2\pi}\int_{0}^{2\pi}\!{\rm d}\phi\;\;
\tau_{\rm host}\!\left(\vec{r}_{\rm p}^{\,\rm sat} + r_{\rm p}\,\hat{n}(\phi)\right),
\end{equation}
where $r_{\rm p}$ is the projected radius from the satellite position.

We can decompose this into a spherically symmetric (or isotropic) and an anisotropic component,
\begin{equation}
\tau_{\rm sat}(r_{\rm p}) = \tau_{\rm sat}^{\rm isotropic}(r_{\rm p})+ \tau_{\rm sat}^{\rm anisotropic}(r_{\rm p}).
\end{equation}
The isotropic component is computed by averaging over the azimuthal position of the satellite within the host,
\begin{equation}\label{eq:sat_ksz_iso}
\begin{aligned}
\tau_{\rm sat}^{\rm isotropic}(r_{\rm p})
&= \frac{1}{(2\pi)^2}\!\int_{0}^{2\pi}\!\!{\rm d}\phi\;
\Bigg[\int_0^{2\pi} d\theta\tau_{\rm host}\!\left(r_{\rm p}^{\rm sat}\hat{n}(\theta)
+ r_{\rm p}\hat{n}(\phi)\right) \Bigg]\\
&= \frac{1}{2\pi}\int_{0}^{2\pi}\!{\rm d}\psi\;\,
\bar{\tau}_{\rm host}\!\left(\sqrt{\left(r_{\rm p}^{\rm sat}\right)^{2} + r_{\rm p}^{2}
+ 2\,r_{\rm p}^{\rm sat} r_{\rm p}\cos\psi}\,\right),
\end{aligned}
\end{equation}
where $\psi \equiv \theta - \phi$, so that the argument of $\bar{\tau}_{\rm host}$ is the projected distance from the \emph{host} centre of a point lying at distance $r_{\rm p}$ from the satellite. The anisotropic contribution is the remainder,
\begin{equation}\label{eq:sat_ksz_aniso}
\tau_{\rm sat}^{\rm anisotropic}(r_{\rm p}) \equiv
\tau_{\rm sat}(r_{\rm p}) - \tau_{\rm sat}^{\rm isotropic}(r_{\rm p}).
\end{equation}
Note that the anisotropic contribution is non-zero only if: 1) the host optical depth profile is anisotropic, 2) the satellite distribution within the host is anisotropic, and 3) the two anisotropies are correlated; in other words, the satellite anisotropy must trace the underlying anisotropic gas distribution.

\begin{figure}
    \centering
    \includegraphics[width=\linewidth]{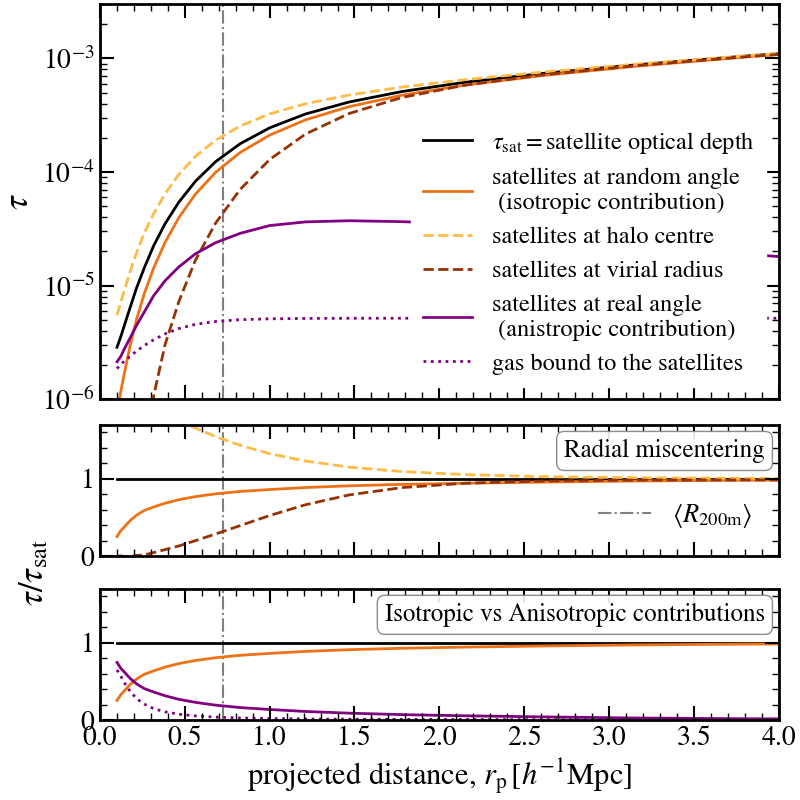}
    \caption{Decomposition of the mean optical depth around satellite galaxies of our DESI LRG-like sample (Table \ref{tab:sample_properties}). The total signal (black) is decomposed into contributions from the isotropic (radial, orange) and anisotropic (azimuthal, purple) satellite distributions within the host halo. The top panel shows the optical depth profiles, while the middle and bottom panels show the ratio of each contribution to the total optical depth around satellites. The isotropic component arises from the spherically averaged gas within the host halo evaluated at the radial position of satellites, and dominates the total signal (bottom panel). This strongly depends on the radial distribution of satellites within the host halo (dashed lines): assuming that all satellites are located at the halo centre (virial radius) overestimates (underestimates) the signal by $\approx40\%$ ($\approx60\%$) at the virial radius. The anisotropic component (solid purple) is primarily sourced by the anisotropic gas distribution around the host halo traced by the satellite positions, while the contribution from gas bound to the satellite subhalos (dotted purple) is subdominant.}
    \label{fig:ksz_sat_effects}
\end{figure}

Figure~\ref{fig:ksz_sat_effects} separates these two contributions from the total optical depth around satellite galaxies, $\tau_{\rm sat}$ in black solid lines. In the upper panel we show the profiles, while in the two lower panels we show the ratio to the total satellite signal. In the bottom panel, we compare the isotropic and anisotropic contributions. The isotropic component (orange solid line) dominates the satellite signal and provides a good description for $r \gtrsim R_{\rm 200m}$. At smaller scales, however, the anisotropic component (purple line) becomes increasingly important: it contributes approximately $20\,\%$ of the total satellite signal at the virial radius and, by $0.2\,R_{\rm 200m}$, exceeds the isotropic contribution.

\subsection{Radial miscentering}\label{sec:sat_misc}

The optical depth around satellites depends on their radial distribution within their host halo -- i.e. satellite miscentering. In this section, we quantify this dependence for the satellites in the DESI LRG-like sample.

This is done by evaluating Eq. \ref{eq:sat_ksz_iso} at different radial distributions $r_{\rm p}^{\rm sat}$. Numerically, each satellite is kept within its own host halo, thus the large-scale amplitude of the signal is unchanged by construction (keeping the same halo mass and any potential assembly bias), so that only the effect of miscentering is isolated.

We display the profiles for three radial distributions. First, we place all satellites at the centre of their host halo (light orange, dashed), representing the case of no miscentering. At the opposite extreme, we place them all exactly at the virial radius (dark orange, dashed), representing maximal miscentering. In between, we place each satellite at its true radial distance but with a randomised azimuthal position (solid orange). 

The upper panel of Figure \ref{fig:ksz_sat_effects} shows the profiles and the middle panel their ratio to the profile measured at the true satellite positions (radial and azimuthal). Two trends are apparent. First, the greater the miscentering, the more extended the profile and the larger the radius at which it reaches its plateau. Second, the greater the miscentering, the more the small-scale amplitude is suppressed and the steeper the profile becomes. This is expected: stacking on miscentered positions washes out the high-density central regions of the halo while spreading the signal to larger radii. If satellites sit at the virial radius, the gas associated with the halo is not fully enclosed until several times the virial radius.

The error from ignoring miscentering (placing satellites at the halo centre) is severe: at the virial radius the signal is overestimated by $\approx 40 \%$ with respect to the true measurement (light orange, dashed in the middle panel of Figure \ref{fig:ksz_sat_effects}). The true profile always lies between the no-miscentering (light orange, dashed) and maximal-miscentering (dark orange, dashed) extremes, closer to one or the other depending on the satellite population and its miscentering distribution. This bracketed region can therefore be interpreted as the maximum theoretical uncertainty arising from an unknown miscentering distribution.

\begin{figure*}
    \centering  
    \includegraphics[width=\linewidth]{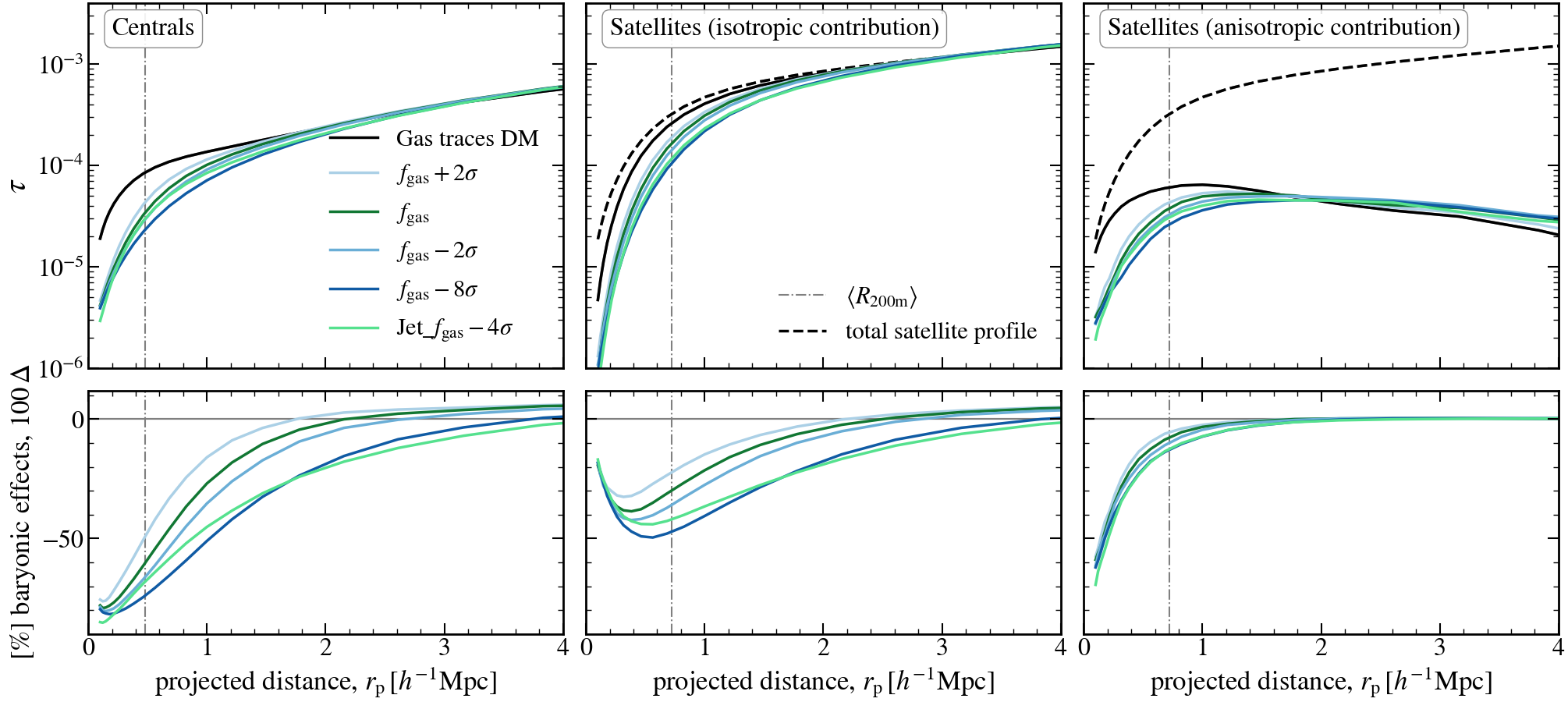} 
    \caption{Impact of baryonic feedback on the mean optical depth around central (left panel) and satellite galaxies on our DESI LRG-like sample (Table \ref{tab:sample_properties}). The latter is decomposed into isotropic (spherically averaged gas profile of the host halo evaluated at the radial position of satellites within the host, middle panel), and anisotropic (extra contribution from the anisotropic distribution of gas around the host halo traced by satellites, right panel) constributions.
    The upper panels show the mean optical depth, and the lower panels show the percentage difference relative to the case in which gas traces the dark matter (black curves). Coloured curves correspond to models with different baryonic feedback strengths, and the vertical dashed lines mark the mean virial radius of the corresponding halo population. Baryonic feedback redistributes gas from halo centres to larger radii, suppressing the optical depth on small scales and producing more extended profiles (left and middle). Around satellites, this suppression is shifted to larger projected radii due to the radial distribution of satellites within their host haloes (middle). The anisotropy of gas is smaller than that of dark matter, in a largely baryon feedback independent way and dominated by hydrodynamical processes such as pressure (right).}
    \label{fig:ksz_baryonic_effect}
\end{figure*}

\subsection{Physical origin of anisotropic component}\label{sec:sat_anisotropy}

The anisotropic component can have two physical origins. First, it may reflect the intrinsically anisotropic distribution of gas around the host halo, which is traced by the satellite population. Although hydrodynamical processes, such as gas pressure, make the gas distribution more isotropic than the underlying dark matter distribution, a residual anisotropy is nevertheless expected as a natural consequence of structure formation. Second, it may arise from gas that remains gravitationally bound to the satellite subhaloes. 

Both contributions are expected to become more important for satellites at larger halo-centric distances beyond the host virial radius, where the large-scale gas and matter distribution are more anisotropic, and recently accreted satellites may not yet have been completely stripped of their gas. In appendix \ref{sec:central_satellite_definition}, we show that indeed this is the case.

To disentangle the origin of this signal, we compute the optical depth from the gas that remains bound to the satellite subhaloes, shown in dotted purple lines in Figure \ref{fig:ksz_sat_effects}. This contribution is subdominant to the total anisotropic component (solid purple), indicating that the anisotropic signal is primarily driven by the anisotropic distribution of gas and satellite galaxies around the host halo, rather than by gas retained within the satellite subhaloes themselves. This result is consistent with \citet{Contreras:2026}, who found that satellite subhaloes are largely stripped of their gas shortly after infall.

In summary, the optical depth around satellite galaxies differs qualitatively from that around centrals due to (1) their miscentering within the host halo, which suppresses the signal on small scales, and (2) their tracing of the anisotropic gas distribution around host haloes, which adds an additional contribution.

\section{Baryonic effects on the optical depth around galaxies}\label{sec:cen_sat_bar}

Unlike dark matter, gas is subject to hydrodynamical and astrophysical processes -- such as star formation, ram pressure and feedback -- in addition to gravity. On the small scales within and around haloes, these processes drive substantial differences between the gas and dark matter distributions, which we collectively refer to as ``baryonic effects''. In this section we show how they affect the optical depth around central and satellite galaxies separately. 

Ideally, baryonic effects would be isolated by comparing a hydrodynamical simulation with its gravity-only counterpart, cross-matching the host haloes and subhaloes of the selected galaxies between the two. This is technically challenging and lies beyond the scope of this paper. However, on the scales of interest, the back-reaction of baryons on the dark matter distribution is small, so we use the dark matter distribution in the hydrodynamical simulation, normalised to the universal baryon fraction, as a proxy for the gravity-only case.

Figure \ref{fig:ksz_baryonic_effect} shows the impact of baryonic physics on the mean optical depth around central galaxies (left column) and satellite galaxies (middle and right columns). For satellites, we further decompose the signal into contributions from the isotropic, spherically averaged, gas and satellite distributions within the host (middle panel) and anisotropic gas and satellite distributions within the host (right panel) (Eq. \ref{eq:sat_ksz_iso} and \ref{eq:sat_ksz_aniso} to see how we compute each of them). 

Different feedback prescriptions in the FLAMINGO simulations are shown by colour: the fiducial model (dark green), and a sequence of weaker (light blue) to stronger (dark blue) feedback models. The kinetic feedback variant is shown by the "Jet" runs, in contrast to the default thermal implementation. The black curves show the case in which gas perfectly traces the dark matter distribution, referred to as dark matter signal for simplicity.

The upper panels of Figure \ref{fig:ksz_baryonic_effect} show the mean optical depth profiles, while the lower panels show the percentage effect of baryons, defined as
\begin{equation}
 \Delta = \frac{\tau_i^{\rm gas} - \tau_i^{\rm dm}}{\tau_j^{\rm dm}}
\end{equation}
where $\tau^{\rm dm}$ denotes the case when gas traces the dark matter. For centrals (first column), $i=j=\text{centrals}$. For satellites (second and third columns), the denominator $j$ is the total satellite optical depth and the numerator, $i$ is either the isotropic or anisotropic component. 

\subsection{Baryonic effects around central galaxies}
In the left column of Figure \ref{fig:ksz_baryonic_effect} we show the impact of baryonic effects on the optical depth around central galaxies. Baryonic feedback evacuates gas from the inner regions of the halo towards the outskirts, lowering the gas density on small scales and making the gas profiles more extended than those of the dark matter.

On large scales, at several virial radii, all the gas evacuated from the centre is recaptured within the aperture, and the gas and dark matter optical depth profiles agree. As expected, stronger feedback produces more extended gas profiles, pushing the gas-dark matter discrepancy out to larger scales.

On small scales, the gas signal is depleted relative to dark matter: at the virial radius the suppression reaches up to $80\%$ in the most extreme feedback scenario, compared with about $60\%$ in the fiducial model. Part of this suppression is expected to come from the conversion of gas into stars. 

While all thermal models exhibit the same scale dependence, the Jet models suppress the gas density more strongly on large scales and less strongly on small scales compared to the thermal models \citep[see also][]{OndaroMallea:2025}. In the future, this difference in scale dependence could be used to distinguish between feedback prescriptions in stacked kSZ analysis and to inform hydrodynamical simulations.

\subsection{Baryonic effects around satellite galaxies}\label{sec:sat_gas_host_th}
\begin{figure*}
    \centering
    \includegraphics[width=\linewidth]{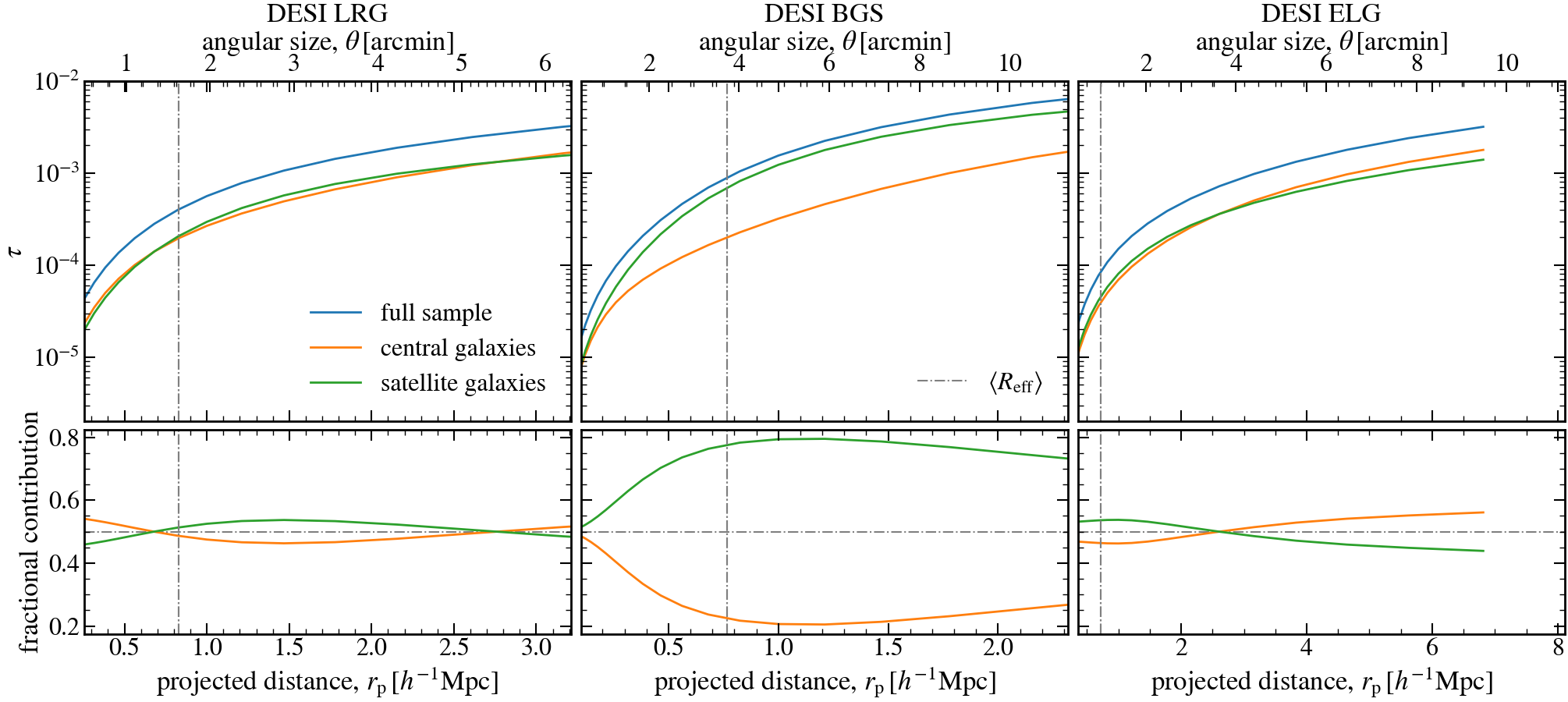}
    \caption{Contribution of stacks around central (orange) and satellite (green) galaxies to the total optical depth (blue) around  DESI-like LRG (left), BGS (middle), and ELG (right) samples. 
    Upper panels show the mean optical depth profiles, while lower panels show the fractional contribution of the central and satellite components to the total signal. The horizontal dashed line marks equal ($50\%$) contributions, and the vertical dashed line the effective virial radius of each sample. Although the satellite fraction in all three samples is considerably smaller than $50\%$, they contribute at least half of the total optical depth (for LRGs and ELGs), and dominate the signal in the BGS sample, accounting for up to $\approx 80\%$ of the signal across all scales.}
    \label{fig:ksz_cen_sat_contr}
\end{figure*}

The middle and right panels of Figure \ref{fig:ksz_baryonic_effect} show the impact of baryonic effects on the optical depth around satellites. The total optical depth around satellites if gas traced the dark matter distribution is shown in dashed black lines, split into the isotropic (solid black, middle panel), and anisotropic contributions (solid black, right panel). We analyse the baryonic effects in each of these components in turn, as they may have distinct physical origins.

\subsubsection{Isotropic component} 
In the middle panel of Figure \ref{fig:ksz_baryonic_effect} we isolate the baryonic effects on the isotropic component -- optical depth from the isotropic, spherically averaged gas within the host evaluated at the satellite radial position, Eq. \ref{eq:sat_ksz_iso} -- relative to the total satellite signal. Qualitatively this is the same process as for centrals, namely gas being evacuated from the halo centre to the outskirts, but with a few differences.

First, because satellite galaxies sit at some radial distance from the halo centre, the strong gas suppression in the central regions of the halo is washed out, and the maximum suppression relative to dark matter is pushed outward, peaking near the virial radius.

Second, the maximum suppression relative to dark matter is smaller for satellites than for centrals. For example, in $f_{\rm gas}-8\sigma$ the amplitude is suppressed by $\approx 60\%$ at twice the virial radius for centrals, but only $\approx 40\%$ for satellites. This is because satellites live in higher-mass haloes than centrals on average. In more massive haloes the deeper potential wells make feedback less effective at evacuating gas, so the suppression relative to dark matter is weaker. 

Third, some of the differences between feedback models may arise from differences in the radial distribution of satellites, rather than solely from differences in the gas distribution. Recent work using the MilleniumTNG hydrodynamical simulations \citep{Parkmor:2023} suggests that the satellite radial positions differ between gravity-only and full physics simulations \citep{Contreras:2026a}: in full-hydrodynamical runs, satellite subhaloes lie closer to the host halo centre than in their dark-matter-only counterparts. Whether this trend is driven primarily by changes in the underlying mass profile (and therefore scales with feedback strength) or by other hydrodynamical processes remains unclear. A definitive answer would require tracking satellites across simulations and comparing their evolution in detail, which is beyond the scope of this work. 

Therefore, we restrict ourselves to comparing the mean miscentering across different feedback variations. Comparing the weakest ($f+2\sigma$) and strongest ($f-8\sigma$) thermal feedback prescriptions, we find that satellites in the weak-feedback calibration lie $\approx 20\,h^{-1}\mathrm{kpc}$ closer to the halo centre than in the strongest-feedback case. 

\subsubsection{Anisotropic component} 
In the right panel of Figure \ref{fig:ksz_baryonic_effect} we isolate the baryonic effects on the anisotropic component -- optical depth from the anistotropic gas and satellite distributions around the host, Eq. \ref{eq:sat_ksz_aniso}--, relative to the total satellite signal. 

First, the anisotropy of the dark matter is larger than that of gas (upper right panel). Second, the difference between gas and dark matter is largely insensitive to the feedback model, remaining below $10\%$ at the virial radius across a wide range of prescriptions (lower right panel). This indicates that the difference between dark matter and gas is driven primarily by hydrodynamical processes, such as ram pressure stripping, rather than by baryonic feedback. 

In summary, the baryonic effects around central and satellites can be divided into two: 1) feedback, which creates an extended gas distribution around the host halo and suppresses the optical depth around both centrals and satellites; and 2) largely feedback-independent hydrodynamical processes, which isotropise the gas distribution around host haloes and suppresses the optical depth around satellites only.

\section{Degeneracy of satellite galaxies and baryonic feedback}\label{sec:sat_bar_deg}

In the previous sections, we studied how stacking on satellite galaxies and baryonic effects change the optical depth profile on small scales. In this section, we show that these two effects are degenerate for realistic galaxy samples, and that the satellite contribution can be misinterpreted as evidence for strong baryonic feedback. 

Quantitatively, the degeneracy depends on (i) the fractional contribution of stacks centred on satellite galaxies to the total stacked signal (Section~\ref{sec:contr_sat}), and (ii) the relative impact of satellites and baryonic feedback to the stacked signal (Section~\ref{sec:deg_bar_sat}). Finally, in Section \ref{sec:toy_analysis_bias}, we perform a mock analysis and show that neglecting the contribution of satellites in the modelling biases the inferred halo gas fractions and feedback strength.

In this section, we use three DESI-like galaxy samples, for which stacked kSZ measurements already exist: LRGs, BGS, and ELGs. The details of the samples are listed in Table \ref{tab:sample_properties}. We do not attempt to reproduce the observed profiles exactly, but rather to capture a representative behaviour. All quantitative comparisons use the covariance of DESI DR2 $\times$ ACT stacks, rescaled to the mock amplitude \citep{Qu:2026,Hadzhiyska:2026}. The figures in this section and the next share a common layout: each column corresponds to a galaxy sample (DESI LRGs, BGS, and ELGs, from left to right), and the error bars show the relative measurement uncertainties of the DESI DR2 stacks from \cite{Qu:2026,Hadzhiyska:2026}. All profiles are convolved with the ACT beam, as described in Section~\ref{sec:estimator}. Accordingly, we define an effective virial radius for each sample that accounts for beam smearing, $r_{\rm eff} = \sqrt{R_{\rm 200m}^2 + \sigma^2}$, where $\sigma = \mathrm{FWHM}/\sqrt{8 \ln 2}$ is the Gaussian width of the beam.

\subsection{Contribution of satellites to the kSZ signal}\label{sec:contr_sat}
The relative importance of satellites in the full sample -- the percentage of the total signal contributed by the stacks around satellite galaxies -- depends mainly on two factors: the satellite fraction of the sample, and the ratio between the mean halo mass of satellites and that of centrals.

The first is straightforward: the larger the satellite fraction in the sample, the greater their contribution to the total signal. The second is less direct: because the signal scales approximately with halo mass, the contribution of satellites -- which preferentially populate massive haloes -- is up-weighted in the average. For the BGS sample, each satellite galaxy contributes roughly 30 times for signal to the sample than each central galaxy, while for LRG and ELG samples each satellite contributes as much as $\sim 6-7$ centrals (Table \ref{tab:sample_properties}). This implies that the satellite fraction alone is not sufficient to quantify the contribution of satellite galaxies to the total kSZ signal, which can be significantly larger.

In Figure~\ref{fig:ksz_cen_sat_contr}, we show the mean optical depth of each DESI-like sample (in columns), split into contributions from centrals (dashed lines) and satellites (dotted lines). The lower panels show the fraction of the total signal contributed by each component.

In all three cases, the satellite contribution peaks around the virial radius of the sample and then decreases slowly at larger radii. Satellites are most important in the BGS sample, where they dominate the signal across \textit{all} scales, contributing up to $\approx 80\%$ of the signal around the virial radius. For the LRG and ELG samples, centrals and satellites contribute roughly equally.

As discussed in Section~\ref{sec:halo_subhalo_finding}, these results are sensitive to the algorithm used for substructure identification and to the criterion used to classify a given substructure as a satellite or a central galaxy. In Appendix~\ref{sec:satellite_definition}, we re-evaluate this analysis using alternative satellite definitions. While the exact values can vary by up to $\approx 20\%$, the qualitative picture remains unchanged across different algorithm choices.

\subsection{Satellites vs baryonic feedback}\label{sec:deg_bar_sat}
\begin{figure*}
    \centering
    \includegraphics[width=\linewidth]{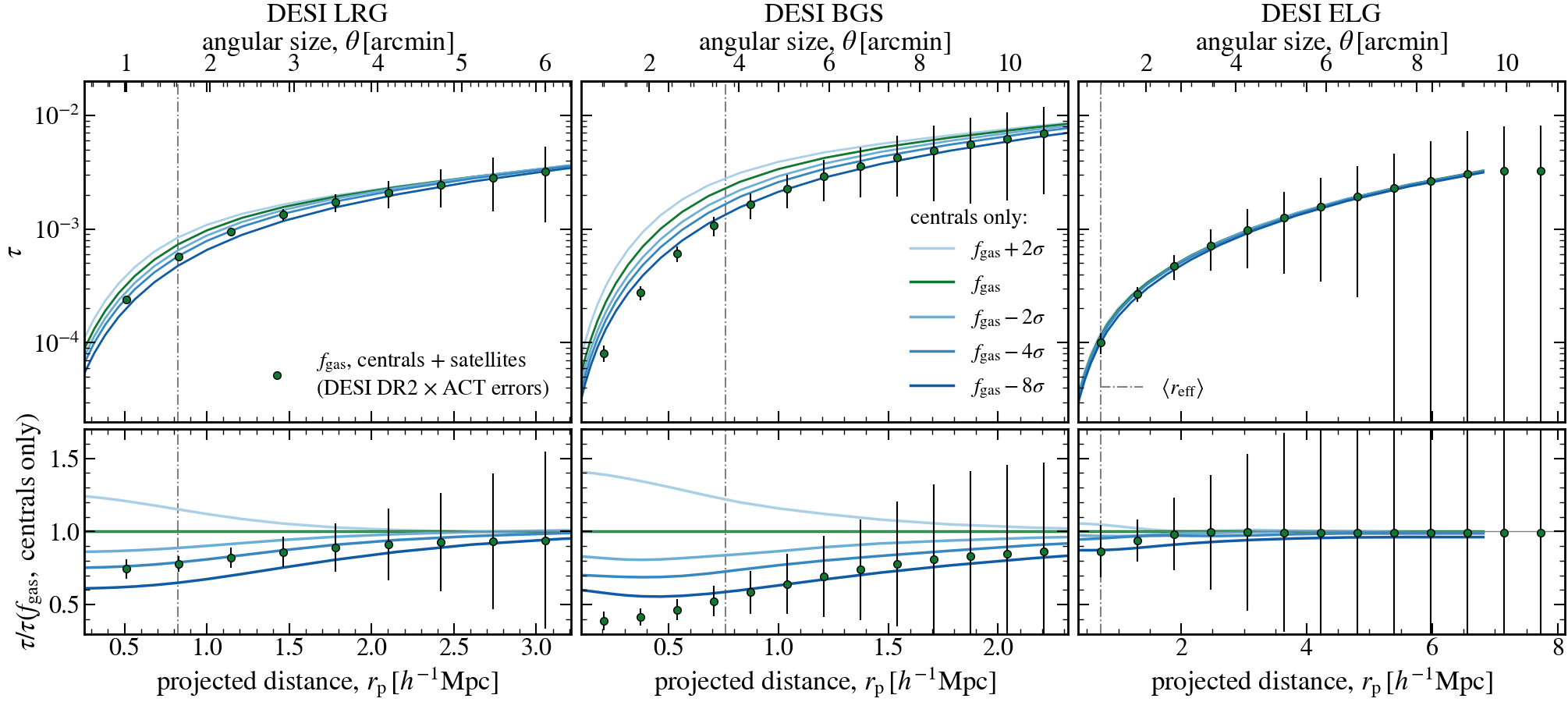}
    \caption{Impact of baryonic feedback and satellite galaxies on the stacked optical depth profiles around DESI LRG, BGS and ELG-like samples. Solid coloured curves show the profiles for varying feedback strengths, stacked on central-only samples. The dotted points show the profile in a fiducial feedback scenario ($f_{\rm gas}$), stacked in galaxy samples including satellites. The associated error bars correspond to the statistical uncertainty of DESI DR2 $\times$ ACT measurements \citep{Qu:2026,Hadzhiyska:2026}. The upper panels show the profiles, and the lower panels the ratio to the fiducial feedback scenario stacked on centrals only. Baryonic feedback and including satellites in the stacking sample impacts the optical depth profiles to a comparable amplitude.}
    \label{fig:baryons_satellites_degeneracy}
\end{figure*}

In Figure \ref{fig:baryons_satellites_degeneracy} we compare the impact of satellite galaxies and baryonic feedback on the stacked optical-depth profile around DESI-like galaxy samples. The solid coloured lines show samples made only of central galaxies\footnote{Including satellites in their correct host halo masses, but at the centre of the host halo (ignoring their miscentering) produces profiles that depart $\approx 10\%$ from the ones fully ignoring satellites.} and different feedback calibrations. The dark green points represent the profiles on the fiducial feedback scenario ($f_{\rm gas}$), introducing satellite galaxies in the sample. Recall that by construction, the mean halo masses of these samples are identical. 

The impact of feedback is similar for the LRG and BGS samples (first and second columns of Figure \ref{fig:baryons_satellites_degeneracy}). 
The extreme feedback scenario ($f_{\rm gas}-8\sigma$) suppresses the signal by $\approx 30\%$ at the effective virial radius relative to the fiducial feedback scenario ($f_{\rm gas}$). For the ELG sample, the effect is milder, with a suppression of $\approx 20\%$ relative to the fiducial case at the virial radius.
\begin{table}
\centering
\begin{tabular}{@{}llr@{}}
\toprule
\textbf{Galaxy sample} &  \textbf{Model} & $\chi^2 / \text{dof}$ \\
 & (centrals only) & \\
\midrule
DESI LRG & $f_{\rm gas}+2\sigma$ & 83.51 \\
 & $f_{\rm gas}$ & 28.49 \\
 & $f_{\rm gas}-2\sigma$ & 6.76 \\
 & $f_{\rm gas}-4\sigma$ & 0.08 \\
 & $f_{\rm gas}-8\sigma$ & 9.96 \\
\addlinespace
DESI BGS & $f_{\rm gas}+2\sigma$ & 484.17 \\
 & $f_{\rm gas}$ & 188.11 \\
 & $f_{\rm gas}-2\sigma$ & 86.46 \\
 & $f_{\rm gas}-4\sigma$ & 41.94 \\
 & $f_{\rm gas}-8\sigma$ & 14.49 \\
\addlinespace
DESI ELG & $f_{\rm gas}+2\sigma$ & 1.19 \\
 & $f_{\rm gas}$ & 0.66 \\
 & $f_{\rm gas}-2\sigma$ & 0.37 \\
 & $f_{\rm gas}-4\sigma$ & 0.29 \\
 & $f_{\rm gas}-8\sigma$ & 0.24 \\
\bottomrule
\end{tabular}

\caption{Goodness of fit between the mock data -- optical depth profiles of the fiducial $f_{\rm gas}$ feedback variation, with centrals and satellites in the sample --, and the model -- optical depth profiles of each feedback variation, with centrals-only in the sample --. We compute the $\chi^2 = (m-d)^{\rm T} C^{-1} (m-d)$ with the covariance from DESI DR2 $\times$ ACT measurements \citep{Hadzhiyska:2026,Qu:2026} rescaled onto the amplitude of the mock data.}
\label{tab:chi2}
\end{table}

On the contrary, including satellites suppresses the signal much more in the BGS sample ($\approx 30\%$ at the effective virial radius) than in the LRG sample ($\approx 10\%$), as shown by the green line vs empty points on Figure \ref{fig:baryons_satellites_degeneracy}. This is because the relative contribution of satellites is much larger on the BGS sample than in the LRG sample (see Section \ref{sec:contr_sat}). In the ELG sample, satellites do not noticeably impact the profiles.

Putting everything together, feedback and satellite galaxies affect the optical depth profiles by similar amounts. The statistical precision of current measurements is already sufficient to distinguish between these scenarios, except for ELGs. Thus, a visual comparison suggests that neglecting satellite galaxies in the modelling can bias the inferred strength of baryonic feedback.

This is quantified in Table \ref{tab:chi2}, where we show the $\chi^2$ values of each feedback model (only centrals) compared to the mock data points (fiducial feedback, centrals and satellites). The $\chi^2$ values are computed taking into account the full covariance (including the correlation between different data points) of the measurements \citep{Hadzhiyska:2026,Qu:2026}.

\subsection{Toy analysis}\label{sec:toy_analysis_bias}
\begin{figure*}
    \centering
    \includegraphics[width=\linewidth]{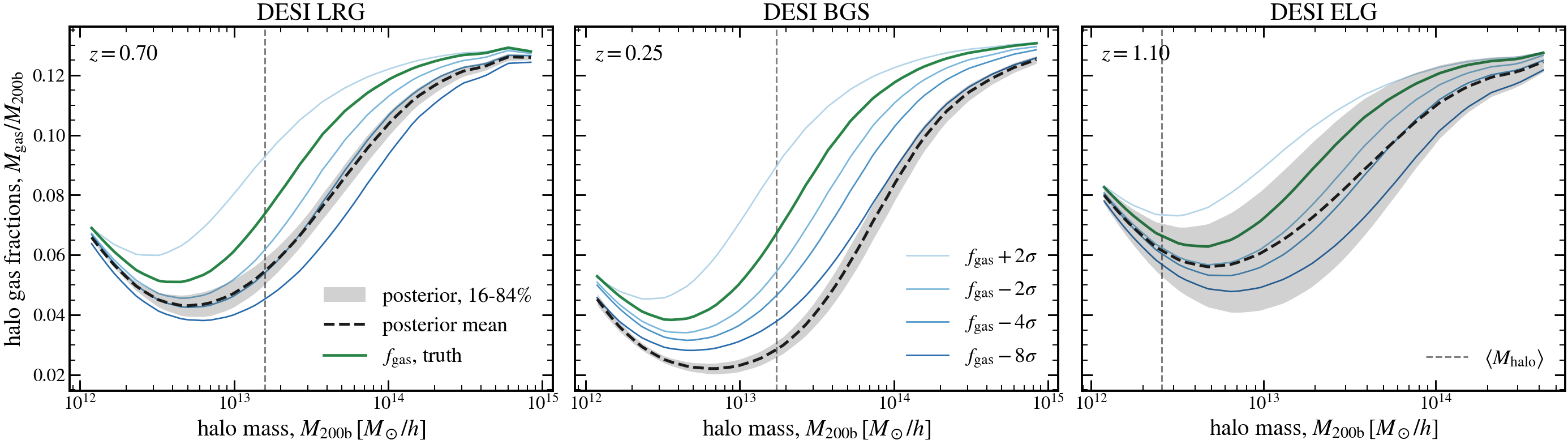}
    \caption{Inferred halo gas fractions obtained by fitting kSZ profiles stacked on centrals and satellites with a model calibrated on centrals only. The gas fractions of the mock data are shown in green, and the inferred ones in black shaded area and dashed line. Thin blue lines show the halo gas fractions of the other FLAMINGO feedback variations for reference. Each column corresponds to a DESI-like sample, with the vertical line indicating its mean halo mass. Ignoring satellites in the modelling biases the inferred gas fractions low, with the largest bias for BGS, followed by LRGs, while no significant bias is found for ELGs due to the large errorbars of the measurements.}
    \label{fig:toy_analysis_bias}
\end{figure*}
In this section we perform a toy analysis of the kSZ profiles stacked on centrals and satellites, using a central-only model, keeping the halo mass of the samples fixed. We find that the inferred halo gas fractions are biased low for DESI BGS and LRG like samples, and unbiased for the DESI ELG-like sample.

For each galaxy sample, we build a Gaussian Process emulator that takes the halo gas fraction as a function of halo mass as input, and learns its relation to the kSZ profiles stacked on centrals only (solid coloured lines in Figure \ref{fig:baryons_satellites_degeneracy}). Details of the emulation and its validation are presented in Appendix \ref{sec:gp_validation}.

We then use the emulator to fit the mock data, consisting of kSZ profiles for the fiducial feedback scenario stacked on centrals and satellites. We run an MCMC with \texttt{emcee}, minimizing $\chi^2 = (m-d)^{\rm T} C^{-1} (m-d)$, where $C$ is the covariance of the measured data \citep{Hadzhiyska:2026,Qu:2026}, rescaled to match the amplitude of the mock data. The best-fit model provides an excellent fit, with $\chi^2 / \text{dof }<1$ for all three samples.

In Figure \ref{fig:toy_analysis_bias}, we show the real (dark green) and inferred halo gas fractions (black dashed lines and shaded area). As expected from the visual comparison in Figure \ref{fig:baryons_satellites_degeneracy}, the BGS sample exhibits the largest bias: the inferred halo gas fractions are approximately half the true values at the mean halo mass of the sample. For LRGs, the impact of satellites on the profiles is smaller than for BGS, resulting in a correspondingly smaller bias: the inferred gas fractions are approximately $40\%$ lower than the true values at the mean halo mass. For the ELG sample, we find no significant bias, owing to the large measurement uncertainties and the relatively small impact of satellites on the profiles.

In summary, satellite galaxies contribute a substantial fraction ($50-80\%$) of the stacked kSZ signal in realistic DESI-like galaxy samples. Relative to central-only samples, including satellites alters the measured profiles at a level comparable to that induced by baryonic feedback. Satellite contamination therefore represents a significant theoretical systematic: neglecting it biases the inferred gas fractions low and leads to an overestimate of the strength of baryonic feedback at the precision of current measurements.

\section{How to account for satellite galaxies}\label{sec:implications}
In this section we discuss several approaches to account for the satellite population on the stacked kSZ profiles, in the context of the statistical precision of current and upcoming measurements. We then state the modelling requirements they imply.

We recall that all the analysis is done at fixed halo mass of the galaxy sample\footnote{The mean halo mass cannot be reliably inferred from kSZ measurements alone, as the large-scale signal has low signal-to-noise and both large- and small-scale profiles are degenerate with other effects (see Section~\ref{sec:background}). Lensing measurements therefore provide a natural complementary probe to calibrate the mass, including galaxy-galaxy lensing measurements \citep{McCarthy:2025b,Siegel:2025} and stacked CMB lensing measurements \citep{Hadzhiyska:2025,Hadzhiyska:2026,Qu:2026} using the same galaxy samples as those employed for kSZ stacks.}, thus the only degree of freedom left in the galaxy sample is its satellite population.

The figures in this section share a common layout, with columns corresponding to different DESI-like samples. Error bars or grey shaded regions show the statistical uncertainties of stacked measurements of ACT maps on DESI DR2 samples \citep{Qu:2026,Hadzhiyska:2026}. The significance of departures from the fiducial values quoted throughout the text is computed using the full covariance matrix of these measurements.

\subsection{In the absence of external datasets: uncertainty from satellite population}\label{sec:sat_marg}
\begin{figure*}
    \centering
    \includegraphics[width=0.8\linewidth]{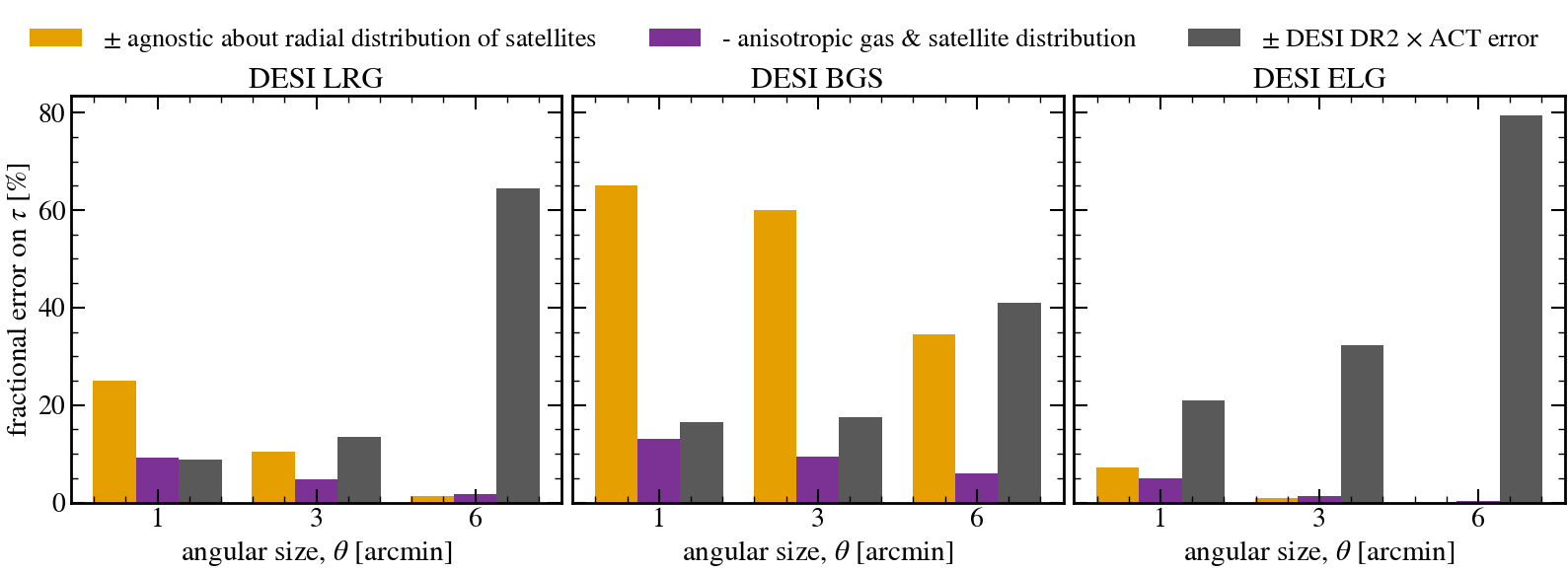}
    \caption{Comparison of theoretical and statistical errors on the stacked kSZ profiles, for DESI LRG-like (left), DESI BGS-like (middle) and DESI ELG-like (right) samples at three angular distances from the galaxy positions. The theoretical errors are computed from a) fully agnostic model about the satellite distribution within the host haloes (orange), and b) the error done by neglecting the contribution from anisotropic gas traced by satellites (purple). The statistical errors come from DESI DR2 $\times$ ACT measurements \citep{Qu:2026, Hadzhiyska:2026} (grey). The dominant theoretical error arises from the unknown radial distribution of satellites within the host, dominating over the statistical uncertainty on small scales for LRG and BGS-like samples. The anisotropic contribution is comparable or subdominant to the current statistical uncertainties ($1-1.5\sigma$ effect), but expected to become significant with upcoming measurements.}
    \label{fig:miscentering_subhalo_errors}
\end{figure*}

In the absence of external datasets constraining the satellite population, its impact on the kSZ profiles is uncertain. The dominant source of this uncertainty is the unknown radial miscentering of satellites with respect to the host halo centre. Here, we quantify the resulting miscentering uncertainty for a fully agnostic model of the satellite population as:
\begin{equation}\label{eq:misc_unc}
 \Delta \tau_{\rm miscentering} (r_{\rm p})= \tau_{\rm sat}^{\rm isotropic}(r_{\rm p}; r_{\rm p}^{\rm sat}=0) -  \tau_{\rm sat}^{\rm isotropic}(r_{\rm p}; r_{\rm p}^{\rm sat}=R_{\rm 200m}),
\end{equation}
where $\tau_{\rm sat}^{\rm isotropic}$ is given by Eq. \ref{eq:sat_ksz_iso}. This represents an upper bound on the miscentering uncertainty, bracketing the cases where satellites are in the centre of their host halo ($r_{\rm p}^{\rm sat}=0$), and the case where they are at the virial radius ($r_{\rm p}^{\rm sat}=R_{\rm 200m}$).

\begin{figure}
    \centering
    \includegraphics[width=0.9\linewidth]{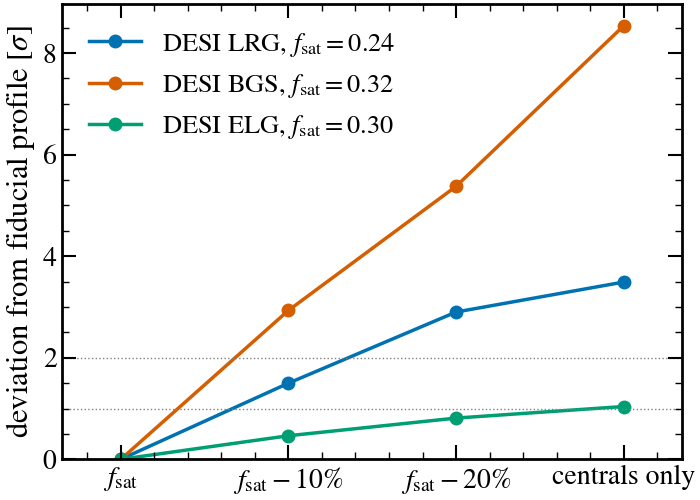}
    \caption{Sensitivity of the stacked kSZ profiles to uncertainties in the satellite fraction for DESI LRG, BGS and ELG-like samples. We assess the impact of uncertainties on the satellite fraction by randomly removing increasing fractions of satellite galaxies while simultaneously removing low-mass central galaxies to keep the mean halo mass of the sample fixed. For each perturbed sample, we measure the deviation of the stacked kSZ profile from the profile obtained using the fiducial satellite fraction. We quantify this deviation as a function of the fractional error in the satellite fraction using the covariance matrix of DESI DR2 $\times$ ACT measurements \citep{Qu:2026,Hadzhiyska:2026}. For the LRG and BGS samples, fractional uncertainties of $\approx10\%$ in the satellite fraction already produce deviations of $\approx1.5\sigma$ and $\approx3\sigma$, respectively, from the fiducial profiles.}
\label{fig:error_fsat}
\end{figure}

In Figure \ref{fig:miscentering_subhalo_errors} we compare the resulting miscentering uncertainties\footnote{Even though the error is not completely symmetric around the fiducial profile, we show it as symmetric for visualization purposes.} (orange) with statistical uncertainties from DESI DR2 $\times$ ACT measurements (grey) in three angular positions: 1, 3, and 6 arcmins from the galaxy positions respectively. 

The result varies significantly across the samples. It is largest for BGS, exceeding 50\% at  $\theta \leq 3 \, \rm arcmin$ and dominating over statistical uncertainties until $\approx 6 \rm \, arcmin$. Marginalizing over an unconstrained satellite distribution would therefore largely erase the BGS sample’s constraining power (recall that even the most extreme feedback scenarios differ by only $\approx 40\%$ at the virial radius; Figure~\ref{fig:baryons_satellites_degeneracy}).

For LRGs, the miscentering uncertainty reaches more than 20\% at $\approx 1 \, \rm arcmin$ and noticeably degrades the constraints when included in the error budget. In contrast, it is $< 10\%$ for ELGs and remains subdominant to their observational uncertainties.

In Appendix~\ref{sec:baryons_errors} we study the dependence of the miscentering uncertainty on baryonic feedback. We find that the absolute uncertainty decreases with stronger feedback, while the relative uncertainty (normalised by the corresponding optical depth) remains consistent across feedback models.

\subsection{Constraining satellite population with external datasets}\label{sec:sat_external_data}

A more informative approach than being fully agnostic about the satellite population is to incorporate complementary observations that directly constrain it. Lensing and galaxy clustering are the most natural candidates, as they constrain both the abundance and radial distribution of satellite galaxies.

A key question is how well the satellite population must be constrained to avoid statistically significant changes in the kSZ profiles. To quantify this, we compare kSZ profiles obtained for slightly different satellite populations, thereby mimicking the effect of an imperfect model for the satellite distribution. Specifically, we artificially reduce the satellite fraction of our mock DESI-like samples while keeping the mean halo mass fixed. Starting from the fiducial sample, we randomly remove a fraction $\Delta f_{\rm sat}$ of the satellite galaxies\footnote{The result depends on how satellites are removed from the sample. For example, replacing only the lowest-stellar-mass satellites with central galaxies produces a smaller effect, as these satellites typically reside in lower-mass haloes and therefore contribute less to the stacked kSZ signal. We argue, however, that randomly removing satellites across the host halo-mass distribution provides a more realistic representation of uncertainties in the theoretical satellite population.} and subsequently remove central galaxies residing in low-mass haloes such that the mean halo mass of the resulting sample remains unchanged.

The results are shown in Figure~\ref{fig:error_fsat}. We show the deviation of the profiles from the fiducial case as a function of the error in the satellite fraction. The significance of the deviation is computed using the covariance of the observational measurements, normalised to the mock profiles. 

As expected, reducing the satellite fraction has the largest impact on the samples most affected by satellite galaxies, namely the LRG and BGS samples. A $10\%$ reduction in the satellite fraction produces statistically significant changes in the BGS profile ($\approx 3\sigma$), while the effect remains at the $\approx 1.5\sigma$ level for LRGs. In contrast, for ELGs, even the extreme assumption that the sample consists entirely of central galaxies remains consistent with the statistical uncertainties.

Alternatively, stacked kSZ and lensing measurements of the same samples can be combined through their ratio to obtain “observational gas fractions”. At first order, this ratio cancels the impact of satellites, avoiding the need to explicitly model the satellite population (see Section \ref{sec:disc_gasfrac}).

\subsection{Higher order effects: contribution from anisotropic gas and satellites}\label{sec:gas_subh}
A secondary, higher-order effect arises from the anisotropic gas distribution around haloes traced by satellite galaxies (Section~\ref{sec:sat_anisotropy}). We quantify this contribution as the difference between the signal centred on a satellite galaxy and the signal evaluated at the same radial distance from the host-halo centre but at a random azimuthal angle (Eq. \ref{eq:sat_ksz_aniso}).

The resulting effect is shown in purple in Figure~\ref{fig:miscentering_subhalo_errors}. It contributes approximately $10\%$ to the signal for LRG and BGS samples at $\theta = 1\, \rm arcmin$. At the precision of current measurements, this is a $1-1.5\sigma$ effect. While it is not yet statistically significant, it is expected to become important with upcoming high signal-to-noise observations, such as those from the Simons Observatory (SO) combined with LSST or Euclid. 

\subsection{Implications for models}\label{sec:models}

The findings of this work impose several requirements on the theoretical models of the stacked kSZ, specifically, which physical effects the gas and satellite modelling should capture.

The dependence of the gas distribution on halo mass can be modelled using hydrodynamical simulations \citep[e.g.][]{Schaye:2015, McCarthy:2017, Pillepich:2018, Dave:2019, Parkmor:2023, Schaye:2023}, or using baryon-painting techniques applied to gravity-only $N$-body simulations \citep[e.g.][]{Arico:2024,Schneider:2025,Osato:2023}. In both approaches, the satellite population can be modelled and constrained using external datasets self-consistently: it is naturally included in hydrodynamical simulations, while in gravity-only simulations it can be introduced through halo-galaxy connection models, such as Halo Occupation Distributions (HODs) \citep{Benson:2000,Peacock:2000,Berlind:2003,Zheng:2005,Guo:2016,Wechsler:2018} or Subhalo Abundance Matching (SHAM) techniques \citep{Vale:2006,Conroy:2006,Contreras:2021}. 

Another option is to model the gas profile analytically \citep[][]{Amodeo:2021,Bigwood:2024,Kovac:2025,Siegel:2026}. In this approach, the contribution of satellite galaxies can be included by convolving the analytic gas density profile with the satellite radial distribution via a HOD (Eq. \ref{eq:sat_ksz_iso}) \citep{Popik:2025,Hadzhiyska:2026,Qu:2026}.

At the current precision, all three approaches -- hydrodynamical simulations; N-body simulations with a baryon painting and a galaxy-halo connection model; analytical gas profile convolved with the satellite radial distribution -- are enough to model the stacked kSZ profile to within $\approx 1-1.5 \sigma$. However, for upcoming measurements the extra component arising from the anisotropic distribution of gas and satellites around haloes will become statistically significant (see Section \ref{sec:gas_subh}). Modelling this contribution require capturing the anisotropic structure of the gas and satellite distribution, and their correlation. This is beyond the reach of simple analytic prescriptions, and will likely require simulation-based approaches. It also favours subhalo abundance matching over of HODs. 

In summary, satellites can be accounted for in several ways: 1) if their distribution is unconstrained, the resulting uncertainty significantly weakens the constraining power; 2) external datasets can instead constrain the satellite fraction, which must be known to better than 10\% to avoid statistically significant changes in the kSZ profiles. At the precision of DESI DR2 $\times$ ACT, both analytical and simulation-based models can adequately describe the kSZ signal, provided that satellite galaxies and their miscentering are accounted for. For future surveys, simulation-based models will likely be needed, as higher-order contributions from the anisotropic gas distribution traced by satellite galaxies will become statistically significant.

\section{Discussion}\label{sec:discussion}




\subsection{Implications for existing stacked kSZ analysis}
Most stacked kSZ analyses in the literature, 
find that the data prefer low gas fractions or strong baryonic feedback. These works adopt different analysis strategies: some are based on analytical models \citep{Amodeo:2021, Bigwood:2024, Kovac:2025, Qu:2026, Hadzhiyska:2026, Siegel:2026}, while others rely on hydrodynamical simulations \citep{Hadzhiyska:2024, McCarthy:2025b,Bigwood:2025,Siegel:2025}; they either include \citep{Hadzhiyska:2024,McCarthy:2025b,Bigwood:2025,Siegel:2025,Qu:2026,Hadzhiyska:2026} or exclude satellites \citep{Amodeo:2021,Bigwood:2024,Kovac:2025,Siegel:2026} in their galaxy sample selections.

Our results are not directly translatable to these analyses. First, many of them focus on stacks around BOSS CMASS and LOWZ galaxy samples, which have significantly lower signal-to-noise than the recent stacks on DESI galaxies on which this work is based (with relative uncertainties at $\theta \approx 1.5-2\, \mathrm{arcmin}$ approximately a factor of $2-2.5$ larger for CMASS galaxies \citep{Schaan:2021} than for DESI DR2 LRGs \citep{Qu:2026}). Consequently, the theoretical error introduced by neglecting satellite galaxies, or by an imperfect satellite calibration, is expected to be less important in those analyses compared to the statistical uncertainties.

Second, our work isolates the impact of satellites by fixing the mean halo mass of the sample. In contrast, these analyses generally vary several parameters simultaneously, in particular the galaxy selection and the mean halo mass of the sample. As a result, the impact of satellite miscentering can be partially absorbed by changes in other parameters rather than directly biasing the inferred gas fractions. Quantifying this is beyond the scope of this work.

\subsection{Including satellites: degeneracy between miscentering and halo mass}\label{sec:fixed_n}
As pointed out in the literature, including satellites in a galaxy sample at fixed number density has two competing effects: it suppresses the central signal through miscentering of satellites relative to the halo centre, while simultaneously increasing the overall amplitude because satellites preferentially reside in more massive haloes than central galaxies of the same stellar mass. These effects partially compensate on small scales, so introducing satellites may have little impact on the small-scale kSZ profile. This has been used to argue that modelling satellites may not be as necessary to analyse the stacked kSZ signal \citep{McCarthy:2025b, Bigwood:2025, Siegel:2026}.

We argue that this modelling assumption can obscure the interpretation of the measured profiles. In particular, analysing a sample containing both central and satellite galaxies with a central-only model generally leads to an underestimate of the mean halo mass of the underlying galaxy population. This appears to be the case in \citet{Siegel:2025} (Table~2).

This has two important consequences. First, it can impact the inferred strength of baryonic feedback, since the quantity of interest -- the halo gas fraction as a function of halo mass -- depends directly on the inferred halo mass.

Second, it can suggest that the impact of satellites is largest on \textit{large} rather than \textit{small scales}. This is the argument made in Appendix D of \cite{Siegel:2026}: a central-only model can adequately describe the small-scale signal of a sample that includes satellites, but cannot simultaneously reproduce the large-scale signal. They therefore conclude that the model can be safely used for small-scale kSZ measurements if the large-scale data points are excluded. We argue, however, that this apparent agreement on small scales results from compensating errors -- underestimating the sample halo mass and neglecting satellites -- that partially cancel on small scales.

\begin{table*}
\centering
\caption{Summary of the findings of this series of papers on interpreting the stacked kSZ effect (this work and \citet{OndaroMallea:2026}). We list the physical origin of the effects that impact the small-scale kSZ profile, the amplitude of the effect, relevance in the context of current and future measurement precision, and the modelling requirements implied by them.}\label{tab:combined_picture}
\begin{tabularx}{\textwidth}{*{5}{>{\centering\arraybackslash}X}@{}}
\toprule
\textbf{Physical origin} &
\textbf{Effect on the measured profile} &
\textbf{Amplitude} &
\textbf{Relevance} &
\textbf{Modelling requirement}\\
\midrule
Satellite miscentering within the host halo & Suppression on small scales &  up to $\approx 50\%$ at the virial radius  & at current precision (DESI DR2 $\times$ ACT) &  both analytical or simulation based \\
\addlinespace
Anisotropic gas and satellite distributions around the halo & Contribution on small scales & $\approx 10\%$ & in future measurements (LSST / Euclid $\times$ SO) & simulation based \\
\addlinespace
Coupling of gas velocity non-linearities to reconstructed velocities & Suppression on small scales &  $\approx 10-20\%$ & in future measurements (LSST / Euclid $\times$ SO) & simulation based\\
\bottomrule
\end{tabularx}
\end{table*}

\subsection{Analysing kSZ profiles vs observational gas fractions}\label{sec:disc_gasfrac}

Analyses that explicitly include satellites typically use some external dataset to calibrate their properties. In the analytical treatment, the HOD models can be informed by the clustering of those galaxy samples \citep{Qu:2026,Hadzhiyska:2026}. Instead, works using hydrodynamic simulations calibrate their galaxy samples so that they reproduce either the measured mean halo mass from CMB lensing \citep{Hadzhiyska:2025}, number density \citep{Hadzhiyska:2024,RiedGuachalla:2025}, or observed galaxy-galaxy lensing signal \citep{McCarthy:2025b,Siegel:2025,Bigwood:2025}. Many of these works then compare the observed kSZ profiles with those produced by the hydrodynamic simulation for the given selection. While this is a promising approach, we caution that the satellite population must be accurately calibrated, as even small miscalibrations can produce statistically significant changes in the resulting kSZ profile (see Section~\ref{sec:sat_external_data}).

Alternatively, \cite{Hadzhiyska:2025} proposed focusing directly on ``observational gas fractions'', obtained by dividing the measured kSZ profile by the measured lensing profile for the same galaxy sample. This observable is more robust to miscalibrations of the satellite population because satellites affect both the kSZ and lensing signals in the same way, causing their impact to largely cancel in the ratio. In light of our results, such strategies are particularly attractive. Following this approach, \cite{Liu:2026} recently found that the fiducial feedback calibration of FLAMINGO most closely tracks kSZ stacks on DESI LRGs and BGSs.

\subsection{Implications for other stacked probes}\label{sec:disc_other_probes}
The same effects apply when other gas probes -- Fast Radio Bursts (FRBs) \citep{McCarthy:2026,Sharma:2026b}, thermal SZ (tSZ) \citep{Liu:2025, Siegel:2026}, and X-ray emission \citep{Popesso:2026} -- are stacked on optically-selected samples that include satellites; particularly, satellite miscentering suppresses the mean signal (unless tSZ or X-ray selected centers are used).

For FRBs, whose signal scales approximately with halo mass as in the kSZ case, our results can be directly translated. However, FRB measurements have substantially less constraining power than kSZ stacks on small scales, so we expect satellite effects to be less statistically important than for kSZ measurements.

For X-ray and tSZ measurements on optically-selected samples and centres, satellite effects are expected to be stronger because the signal has a steeper dependence on halo mass, scaling as $M^{4/3}$ and $M^{5/3}$, respectively. At fixed stellar mass, satellites preferentially inhabit more massive haloes than centrals, further enhancing their contribution to the total signal. Consequently, even a small satellite fraction in the galaxy sample can contribute substantially to the stacked signal, making biases from neglecting satellites or small errors in their calibration potentially larger than those found for kSZ measurements.

We further note that the approach discussed in the previous section to mitigate sensitivity to the satellite population -- analysing the kSZ-to-lensing ratio rather than the kSZ signal alone \citep{Hadzhiyska:2025} -- is likely less robust for tSZ and X-ray measurements. This is because their steeper halo-mass weighting leads to an effective satellite population that differs from that probed by lensing.

\subsection{Further unexplored effects}\label{sec:further}

Several effects not explored in this series of papers may become important for interpreting kSZ measurements as observational precision improves. We briefly discuss three of them.

First, miscentering of central galaxies relative to the minimum of the gas potential may become important for low-redshift samples, such as BGS, and higher-resolution CMB experiments, such as SO. Second, assembly bias in the halo gas content -- systematic variations in gas content at fixed halo mass associated with galaxy properties -- may introduce an additional source of uncertainty. Finally, constraining the stellar distribution on small scales will be particularly important, as a depletion of gas relative to the dark matter profile may result from gas being locked into stars rather than expelled by baryonic feedback.

\section{Summary and conclusions}\label{sec:conclusions}
In this work, we study how satellite galaxies and baryonic feedback jointly shape the stacked kinetic Sunyaev-Zel'dovich (kSZ) signal. They produce degenerate signatures, so that neglecting satellites biases the inferred feedback at the statistical precision of current measurements.

We use the FLAMINGO suite of cosmological-hydrodynamical simulations to construct mock kSZ signals around three realistic DESI-like samples: luminous red galaxies (LRGs), bright galaxy survey (BGS), and emission like galaxies (ELG) (Section \ref{sec:methods}). We fix the mean halo mass of each sample to that inferred from external CMB lensing measurements on the same samples, thereby fixing the large-scale amplitude of the kSZ signal, and isolate the small-scale effects where most of the constraining power lies. 

We summarise our main findings:

\begin{itemize}
 \item The kSZ signal centred on satellite galaxies differs qualitatively from that around centrals (Section \ref{sec:sat_geom}). Because satellites are offset from the centres of their host haloes (\textit{radial miscentering}), stacking the kSZ signal on satellite positions suppresses the average gas profile on small scales. The displaced signal is redistributed to projected radii of order the virial radius, mimicking a genuinely extended gas distribution induced by baryonic feedback (Section \ref{sec:cen_sat_bar}). In addition, satellites trace the anisotropic distribution of gas around the host halo, creating an extra optical depth component at $r\lesssim R_{\rm 200m}$. This component is largely insensitive to the adopted feedback calibration, suggesting that it is governed primarily by structure formation and hydrodynamical processes rather than by feedback.

\item Even a small satellite fraction can contribute as much as, or even dominate, the signal from centrals. This is because the kSZ signal scales with halo mass, while satellites preferentially inhabit more massive host haloes than centrals of the same stellar mass (Section \ref{sec:contr_sat}). 

\item Including satellites in the stack changes the measured profile relative to a central-only stack, with an amplitude comparable to that induced by baryonic feedback (Section \ref{sec:deg_bar_sat}). Consequently, neglecting satellites when modelling the stacked kSZ signal biases the inferred halo gas fractions low and leads to an overestimate of the strength of baryonic feedback (Section \ref{sec:toy_analysis_bias}).

\item Robust constraints on baryonic feedback require satellites to be carefully accounted for. In the absence of external datasets, a fully agnostic satellite model can marginalizing over satellite uncertainties, at the cost of substantially reduced constraining power (Section \ref{sec:sat_marg}). If external data such as lensing and clustering are used instead, they must constrain the satellite population to better than $\approx10\%$ to avoid statistical significant changes in the kSZ profiles (Section \ref{sec:sat_external_data}).

\item These results place requirements on the modelling approaches (Section \ref{sec:models}). At the precision of current data, both simulation-based and analytical models can capture the relevant effects, provided they account for the radial distribution of satellites within host haloes. As upcoming data become more precise, the contribution from the anisotropic distribution of gas and satellites will become statistically significant (Section \ref{sec:gas_subh}), likely requiring simulation-based approaches.
\end{itemize}

In summary, satellite galaxies are an intrinsic component of the stacked kSZ signal, rather than a secondary correction. Ignoring their contribution can bias the inferred gas distribution and halo gas fractions, making the treatment of satellites -- and, more broadly, the characterization of the galaxy sample -- one of the dominant theoretical systematics in current and upcoming stacked kSZ analyses. External datasets, such as CMB and galaxy-galaxy lensing, and galaxy clustering, can help constrain the galaxy-halo connection and satellite distribution, either through forward modelling or through observables that are less sensitive to these effects, such as kSZ–lensing ratios. More broadly, combining kSZ measurements with these complementary datasets can break key degeneracies and enable a robust interpretation of the kSZ signal in terms of the underlying gas distribution and baryonic feedback.

Combining the findings of this paper and its companion paper \cite{OndaroMallea:2026}, we conclude that several physical effects beyond baryonic feedback shape the small-scale profile of the stacked kSZ signal (a summary in Table \ref{tab:combined_picture}). These include radial miscentering of satellites within their host haloes, which is relevant at the precision of DESI DR2 $\times$ ACT measurements, as well as higher-order effects such as the anisotropic gas distribution traced by satellites, and the coupling between nonlinearities in the gas velocity field and reconstructed velocities, which are expected to become relevant for Euclid / LSST $\times$ SO. Fully exploiting current, and particularly future, stacked kSZ measurements to constrain baryonic feedback will require a consistent modelling framework, likely simulation-based, that captures these effects. 

\section*{Acknowledgements}
We thank Simone Ferraro and Giovanni Aricò for helpful discussions during the preparation of this manuscript, and Ian McCarthy and Joop Schaye for their comments on the manuscript. This work was performed in part at the Aspen Center for Physics, which is supported by a grant from the Simons Foundation (1161654, Troyer). REA received support from grant PID2024-161003NB-I00 funded by MICIU/AEI/10.13039/501100011033 and by ERDF/EU. \textit{Author contributions}: \textbf{LOM}: conceptualization; formal analysis; methodology; investigation; visualization; validation; writing-original draft; \textbf{BH}: conceptualization, formal analysis, writing-review\&editing; \textbf{REA}: formal analysis, writing-review\&editing; 

\appendix
\section{Dependence on subhalo definition}\label{sec:satellite_definition}

In this appendix, we assess the dependence of our results on the subhalo-finding algorithm (Section~\ref{sec:subhalo_algorithm}) and on the definition of central and satellite galaxies (Section~\ref{sec:central_satellite_definition}). We find that our conclusions are qualitatively robust to these choices, with quantitative differences below the $\approx10\%$ level in the DESI LRG-like sample (Table \ref{tab:sample_properties}).

\begin{table}
\centering
\begin{tabular}{@{}lrrrr@{}}
\toprule
\textbf{Satellite definition} & $f_{\rm sat}$ & kSZ from sat. & miscentering & anisotropic \\
 & & at $\langle R_{\rm eff}\rangle$ & at $\theta=1'$ & at $\theta=1'$ \\
\midrule
HBT-HERONS & 0.24 & 51\% & 25\% & 9\% \\
VelociRaptor & 0.20 & 40\% & 19\% & 10\% \\
HBT-HERONS, \\$r<R_{\rm 200m}$ & 0.16 & 42\% & 19\% & 3\% \\
\bottomrule
\end{tabular}
\caption{Satellites in the DESI LRG-like sample in the $f_{\rm gas}-8\sigma$ feedback variation, for each substructure finder and satellite definition. $f_{\rm sat}$ is the satellite fraction of the galaxy sample; the third column is the fraction of the total kSZ signal coming from stacks around satellites at $\langle R_{\rm eff}\rangle$; the last two are the theoretical uncertainties at $\theta = 1\,$arcmin, both as fractions of the true satellite profile: the $\pm$ spread by being agnostic to the radial distribution of satellites within the host halo (Eq. \ref{eq:misc_unc}), and the contribution of anisotropic gas traced by satellites (Eq. \ref{eq:sat_ksz_aniso}).}
\label{tab:summary_satellites}
\end{table}

\begin{table}
\centering
\begin{tabular}{llrr@{}}
\toprule
\textbf{Model} & \multicolumn{3}{c}{$\chi^2$} \\
\cmidrule(l){2-4}
(satellites at host) & \textbf{HBT-HERONS} & \textbf{VelociRaptor} & \textbf{HBT-HERONS}, \\ & & & $r<R_{\rm 200m}$ \\
\midrule
$f_{\rm gas}+2\sigma$ & 38.07 & 29.60 & 19.72 \\
$f_{\rm gas}$ & 12.24 & 7.18 & 3.57 \\
$f_{\rm gas}-2\sigma$ & 1.41 & 0.16 & 0.18 \\
$f_{\rm gas}-8\sigma$ & 10.32 & 16.01 & 18.74 \\
\bottomrule
\end{tabular}
\caption{Goodness of fit of DESI LRG-like kSZ profile stacked with the satellites at their true positions, in the fiducial feedback variation ($f_{\rm gas}$). The model consists of miscentering-free (satellites stacked at their host position) profile of each feedback variation. Columns represent different substructure finders and satellite definitions. The $\chi^2 = (m-d)^{\rm T} C^{-1} (m-d)$ is computed with DESI DR2 $\times$ ACT measurement covariances, rescaled onto the amplitude of the mock data's profile. In all cases, $f_{\rm gas}-2\sigma$ feedback is preferred over the fiducial feedback variation of the real data, if satellites are assumed to be at the host halo centre.}
\label{tab:chi2_satellite_definition}
\end{table}

\subsection{Substructure finding: \texttt{HBT-HERONS} vs \texttt{VELOCIraptor}}\label{sec:subhalo_algorithm}
Finding haloes and subhaloes in cosmological simulations is a fundamental step on their analysis. 
Different structure-finding algorithms adopt different strategies, leading to differences in the identified subhalo populations and their inferred properties that depend on simulation resolution, the inclusion of baryonic physics, and other numerical details \citep[see, e.g.,][]{Forouhar:2025}. Here, we investigate how the choice of subhalo finder affects the inferred satellite population and the resulting stacked kSZ profiles by comparing galaxy catalogues constructed with \texttt{HBT-HERONS} and \texttt{VELOCIraptor}. We focus on the $f_{\rm gas}-8\sigma$ feedback variation at the fiducial resolution (\texttt{$\rm L1\_m9$}).

\texttt{HBT-HERONS} is a history-based subhalo finder that follows the evolution of subhaloes across simulation snapshots \citep{Forouhar:2025}. In contrast, \texttt{VELOCIraptor} identifies substructures in phase space by searching for overdensities in both position and velocity, exploiting the fact that subhaloes are kinematically colder than the surrounding background \citep{Elahi:2019}. Using FLAMINGO simulations, \citet{Forouhar:2025} found that \texttt{VELOCIraptor} identifies $\approx 10\%$ fewer satellites than \texttt{HBT-HERONS} around the virial radius, largely independent of halo mass. The differences become more pronounced in the inner regions of haloes, where \texttt{VELOCIraptor} misses a significant fraction of the substructures identified by \texttt{HBT-HERONS}, particularly for host haloes with $M_{\rm 200c}\approx 10^{13}\,h^{-1}\mathrm{M}_\odot$.

In Table \ref{tab:summary_satellites} we summarise the impact of these differences on the satellite population and stacked kSZ signal.
Consistent within \citet{Forouhar:2025}, the satellite fraction in the \texttt{VELOCIraptor} catalogue is lower than in the \texttt{HBT-HERONS} catalogue ($f_{\rm sat}=0.24$ vs $f_{\rm sat}= 0.2$). Consequently, the fractional contribution of the stacks around satelites to the total signal decreases from $51\%$ to $40\%$ at the effective virial radius. Likewise, the theoretical uncertainty associated with the unconstrained miscentering distribution of satellite galaxies reduces from $25\%$ to $19\%$ at $\theta=1\, \rm arcmin$. The contribution of the anisotropic gas distribution traced by satellites remains at the $10\%$ level at $\theta=1\, \rm arcmin$. 

As the relevance of satellites does not dramatically change between \texttt{HBT-HERONS} and \texttt{VELOCIraptor}, we expect that their impact in the kSZ profiles, and therefore, the modelling error due to ignoring them not to change qualitatively either. In Table \ref{tab:chi2_satellite_definition} we repeat the analysis of Table \ref{tab:chi2}, for different subhalo finders\footnote{Unlike in Table \ref{tab:chi2}, as a model we use a sample where satellites are artificially positioned at the centre of the host halo, rather than a central-only sample. Moreover, we do not include the $f_{\rm gas}-4\sigma$ simulation in the comparison. Both differences are because we do not have the central-only sample and $f_{\rm gas}-4\sigma$ run with \texttt{VELOCIraptor} catalogues.}. While the $\chi^2$ values slightly change between \texttt{HBT-HERONS} and \texttt{VELOCIraptor}, these changes are marginal and the qualitative picture is the same for both substructure finders.

\subsection{Satellites within and beyond the virial radius}\label{sec:central_satellite_definition}
Once subhaloes have been identified, they are assigned to haloes defined by a three-dimensional Friends-of-Friends (\texttt{FoF}) algorithm. Within each \texttt{FoF} halo, the most massive subhalo and the galaxy within it are classified as the central, while all remaining subhaloes and galaxies are classified as satellites.

We find that a non-negligible fraction of these satellites lie beyond the virial radius of their host halo. Some of these are genuine satellites, while others are a well-known consequence of the \texttt{FoF} algorithm, which can link neighbouring structures that have not yet merged into a single halo. To obtain a conservative estimate of the impact of this effect, we reclassify all satellites located outside the virial radius of their host halo as central galaxies.

We show the results in Tables \ref{tab:summary_satellites} and \ref{tab:chi2_satellite_definition}. The changes with respect to our fiducial analysis (\texttt{HBT-HERONS}) are modest, remaining below the $\approx10\%$ level in most cases. The exception is the relevance of the anisotropic gas and satellite distribution: at $\theta=1\,\rm arcmin$ falls to $3\%$ if satellites within the virial radius of the host halo are retained, as opposed to a $9\%$ if all satellites are considered (Table \ref{tab:chi2_satellite_definition}). This is expected, as the anisotropy of the gas distribution, a natural consequence of the structure formation, is expected to be larger in the outskirts of haloes. 

Overall, we conclude that our main results are robust to the choice of subhalo-finding algorithm and central-satellite definition. 

\section{Emulation of the stacked kSZ profiles}\label{sec:gp_validation}
The toy analysis of Section~\ref{sec:toy_analysis_bias} needs a model that predicts a stacked kSZ profile for an arbitrary halo gas fraction. We build it as a Gaussian process (GP) trained on the kSZ profiles stacked on central-only samples in 5 feedback variations. We have one emulator per DESI-like sample. In this appendix we describe how it is built (Section~\ref{sec:gp_architecture}) and validate it (Section~\ref{sec:gp_tests}).

\subsection{Architecture and training}\label{sec:gp_architecture}
\begin{figure}
    \centering
    \includegraphics[width=0.9\linewidth]{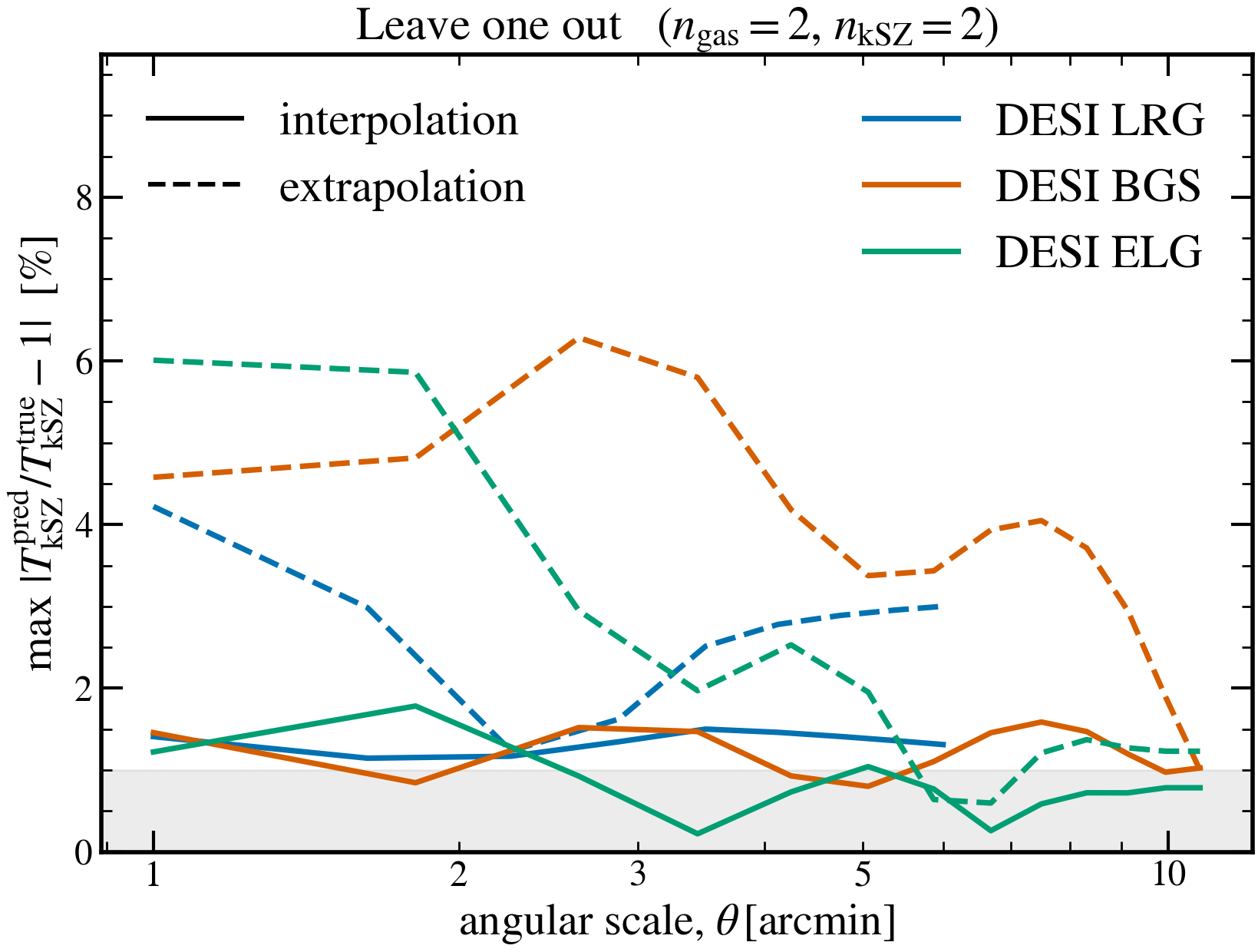}
    \caption{Accuracy of the emulator assessed with a leave-one-out test. Each feedback variation is removed from the training set in turn, and the emulator is rebuilt from the remaining variations. We plot the predicted kSZ profile for the held-out variation, given its halo gas fraction, divided by the true kSZ profile. Curves show the largest fractional error across feedback variations at each angular scale, for each DESI-like sample (colours). Solid lines correspond to interpolation, where the held-out run is bracketed by the remaining training set. Dashed lines correspond to the two ends of the feedback ladder, $f_{\rm gas}+2\sigma$ and $f_{\rm gas}-8\sigma$, which are extrapolations when held out. The interpolation is accurate at the $1-2\%$ level, while the extrapolation is below the $6\%$ in all angular scales.}
    \label{fig:gp_forward_loo}
\end{figure}

\begin{figure}
    \centering
    \includegraphics[width=0.9\linewidth]{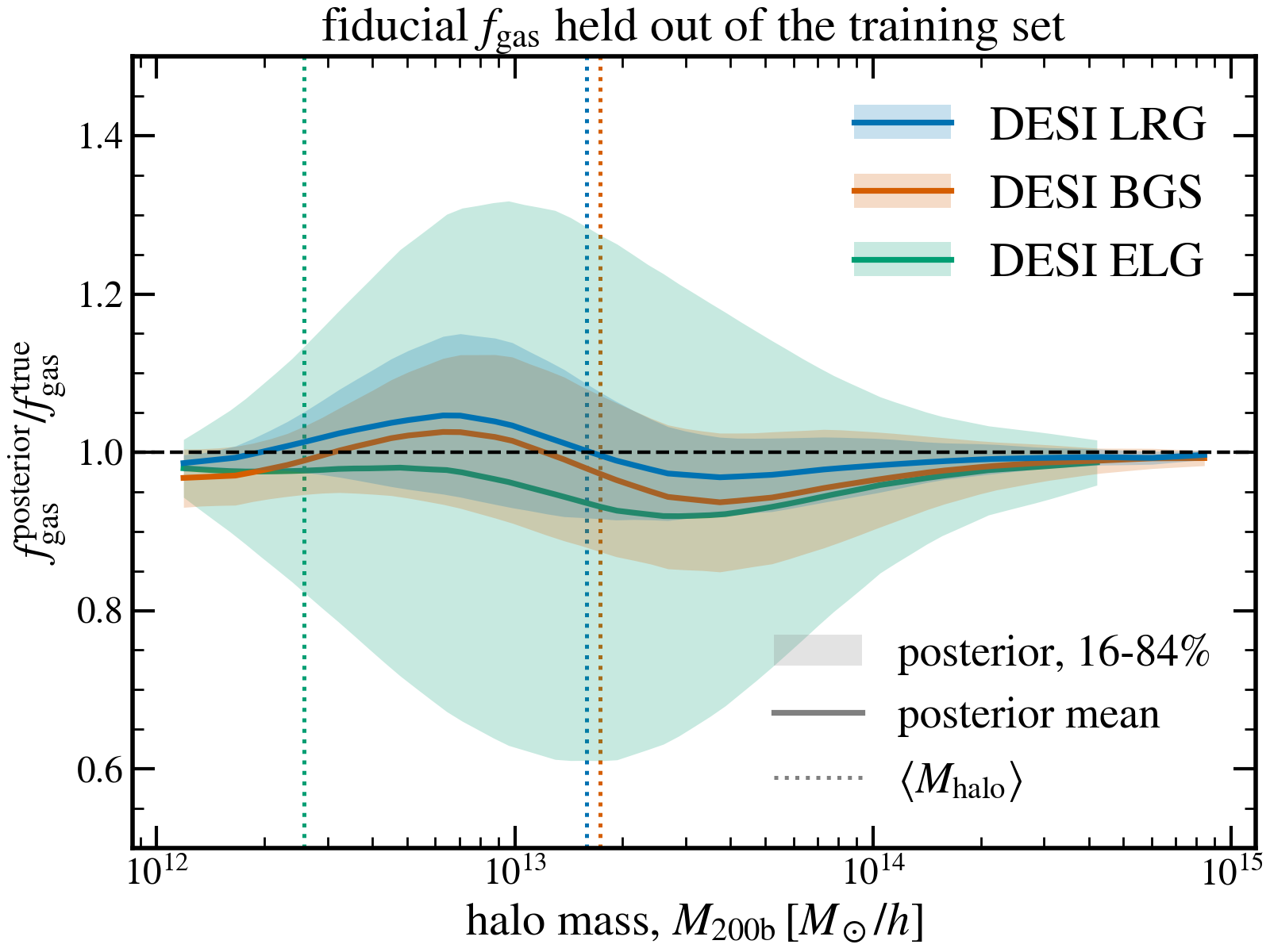}
    \caption{End-to-end test of the emulator. The fiducial feedback variation is held out from the training set, and its kSZ profile is fitted to infer the halo gas fractions. We show the ratio of the posterior halo gas fractions to the real halo gas fractions. Colours correspond to the DESI LRG-like, BGS-like, and ELG-like samples, and the vertical dotted line is the mean halo mass of each. In all samples, the true halo gas fractions are recovered within the $1\sigma$ posterior.}
    \label{fig:gp_holdout}
\end{figure}

The emulator learns a single map,
\begin{equation}
    f_{\rm gas}(M) \;\longrightarrow\; T_{\rm kSZ}(\theta),
\end{equation}
from five training points: the feedback variations variations $f_{\rm gas}+2\sigma$, fiducial $f_{\rm gas}$, $f_{\rm gas}-2\sigma$, $f_{\rm gas}-4\sigma$, and $f_{\rm gas}-8\sigma$ (see Section \ref{sec:flamingo_sims}). Each training point pairs the halo gas fraction, measured in logarithmically spaced mass bins between $10^{12}$ and $10^{15}\, h^{-1} \rm M_\odot$, with its kSZ profile stacked on a centrals-only sample, evaluated at the $\theta$ points of the DESI DR2 $\times$ ACT measurements. The measurements of both the halo gas fractions and kSZ profiles are done at the redshift of the sample (Table \ref{tab:sample_properties}).

Five simulations cannot constrain $f_{\rm gas}(M)$ and $T_{\rm kSZ}(\theta)$ in
every mass and angular bin, so both sides are compressed into principal
components before the GP learns the mapping between them. We find that two
principal components per function are enough to describe both the halo gas
fractions ($\mathbf{c}$, $n_{\rm gas}=2$) and the kSZ profiles ($\mathbf{a}$, $n_{\rm kSZ}=2$) at the $1\%$ level.
The GP therefore learns a map between two pairs of numbers,
\begin{equation}
    f_{\rm gas}(M) \xrightarrow{\ \rm PCA\ } \mathbf{c}
    \xrightarrow{\ \rm GP\ } \mathbf{a}
    \xrightarrow{\ \rm PCA^{-1}\ } T_{\rm kSZ}(\theta) .
\end{equation}
The relation between the two is close to linear -- more gas, more signal -- so
we first fit a linear trend $\mathbf{a} \simeq \beta_0 + \boldsymbol{\beta}^{\top}\mathbf{c}$ by least squares, and the GP learns only the residual
$\tilde{\mathbf{a}} = \mathbf{a} - (\beta_0 + \boldsymbol{\beta}^{\top}\mathbf{c})$.

The GP uses a squared-exponential kernel with a single length scale $\ell$, which
depends only on the distance between the gas fractions,
\begin{equation}
    K_{ij} \;=\; \exp\!\left( -\,\frac{\lVert \mathbf{c}_i - \mathbf{c}_j \rVert^{2}}{2\ell^{2}} \right)\;+\; 10^{-10}\,\delta_{ij} ,
    \label{eq:gp_kernel}
\end{equation}
where $\delta_{ij}$ is the Kronecker delta and the second term is a numerical
jitter added for the stability of the matrix inversion. We include no noise term,
so the emulator reproduces each of its five training runs exactly. We fit for $\ell$ by minimising
\begin{equation}
    -\log \mathcal{L}(\ell) \;=\; \frac{1}{2}\sum_{d=1}^{n_{\rm kSZ}}
    \tilde{\mathbf{a}}_d^{\top}\,K(\ell)^{-1}\,\tilde{\mathbf{a}}_d
    \;+\; \frac{n_{\rm kSZ}}{2}\,\log \bigl| K(\ell) \bigr| .
    \label{eq:gp_nll}
\end{equation}
The two terms pull in opposite directions: the first favours a short $\ell$, for which each simulation is informative mainly about itself and residuals can be accommodated, while the second penalises this flexibility and favours a long $\ell$, for which the five runs are strongly correlated.

Once $\ell$ is fixed, the prediction of the kSZ profile given a new gas fraction is analytic. We compress the new gas fraction into its principal component $\mathbf{c}_*$, evaluate its similarity to each training run,
$k_{*i} = \exp(-\lVert \mathbf{c}_* - \mathbf{c}_i \rVert^{2} / 2\ell^{2})$, and obtain
\begin{equation}
    \mathbf{a}_* \;=\; \beta_0 + \boldsymbol{\beta}^{\top}\mathbf{c}_*
    \;+\; \mathbf{k}_*^{\top} K^{-1} \tilde{\mathbf{a}} ,
    \label{eq:gp_predict}
\end{equation}
which we then decode back into $T_{\rm kSZ}(\theta)$.

\subsection{Validation}\label{sec:gp_tests}

We test the emulator using a leave-one-out approach. Figure~\ref{fig:gp_forward_loo} shows the emulation accuracy. We leave out each feedback variation in turn, rebuild the emulator using the remaining four, and input the gas fraction of the held-out variation. We then compute the ratio between the predicted and true kSZ profiles and, at each angular scale, show the largest error across all feedback variations. When the held-out variation is bracketed by the remaining training set (i.e., the emulator interpolates; solid lines), the profile is recovered to better than $2\%$. When either extreme, $f_{\rm gas}+2\sigma$ or $f_{\rm gas}-8\sigma$, is left out, the emulator instead extrapolates, resulting in a slightly larger error of $\approx 6\%$ (dashed lines). Both errors are smaller than the statistical uncertainties of the DESI DR2 $\times$ ACT measurements used in our analysis and are therefore negligible for our purposes.

In Figure~\ref{fig:gp_holdout}, we test whether fitting the kSZ profiles recovers the halo gas fractions within the posterior uncertainties. We leave the fiducial feedback variation out of the training set, fit its kSZ profiles, and compare the inferred gas fraction posteriors with the true values, following the approach of Section~\ref{sec:toy_analysis_bias}. We perform the fit with an MCMC, minimizing the $\chi^2$ between the data and emulator predictions with the covariance of the DESI DR2 $\times$ ACT measurements. The true gas fraction is recovered within the $16-84\%$ posterior interval for all three DESI-like samples.

\section{Baryonic effects on the miscentering uncertainty}\label{sec:baryons_errors}
\begin{figure}
    \centering
    \includegraphics[width=0.9\linewidth]{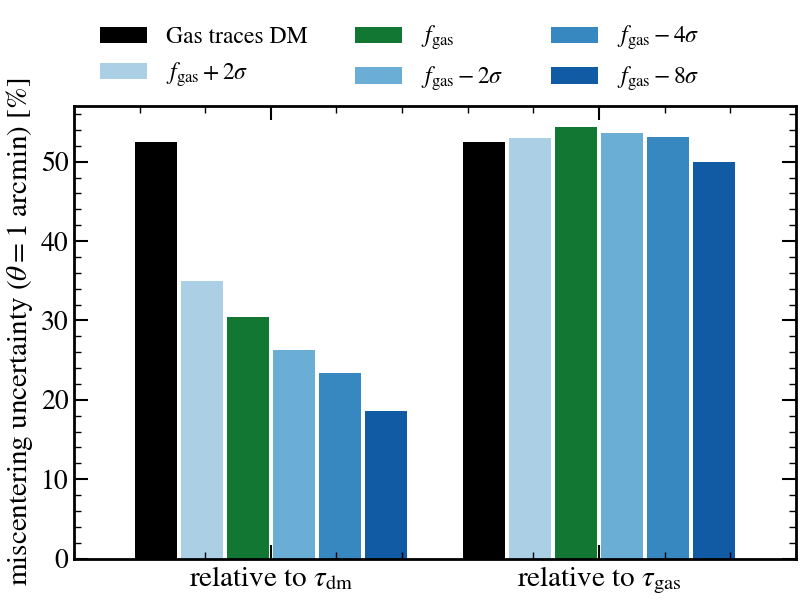}
    \caption{Dependence of the miscentering uncertainty on the underlying density profile, as given by different baryonic feedback models (colours). The uncertainty is quantified by Eq.~\ref{eq:misc_unc} and corresponds to the limiting cases in which all satellites are located either at the centres of their host haloes or at their virial radii. The absolute uncertainty, normalised to the same dark matter optical depth profile, $\tau_{\rm dm}$, decreases with increasing baryonic feedback (left). The relative uncertainty, normalised to the optical depth profile of each corresponding density distribution, $\tau_{\rm gas}$, is nearly independent of the underlying density profile (right).}
    \label{fig:satellite_errors_baryons}
\end{figure}

In this appendix, we investigate the dependence of the miscentering uncertainty (Section \ref{sec:sat_marg}) on baryonic feedback. Figure~\ref{fig:satellite_errors_baryons} shows the miscentering uncertainty of an unconstrained satellite distribution, bracketed by all satellites being located either at the centres of their host haloes or at their virial radii (Eq.~\ref{eq:misc_unc}). The coloured histograms show the uncertainty in several feedback variations, while the black histograms represent the case where gas traces the dark matter distribution. We display the uncertainties at $\theta=1 \, \rm arcmin$ for the DESI LRG-like sample, but have checked that the picture holds for other samples and angular distances.

In the left, we normalise all the uncertainties by the optical depth profile of gas tracing dark matter. As expected, the \textit{absolute} uncertainty is larger for the more centrally concentrated dark matter profile, reaching $\approx 50\%$ of the signal at $\theta=1\,\rm arcmin$. This is because, for a more concentrated profile, the difference between the density at the centre (no miscentering) and at a fixed distance (miscentering) is larger than for a less concentrated profile. For the most extreme feedback scenario considered, $f_{\rm gas}-8\sigma$, the miscentering uncertainty is a factor of $\approx 2.5$ smaller than in the dark matter case at $\theta = 1 \, \rm arcmin$. Thus, in \textit{absolute} terms, the impact of miscentering depends on the underlying density profile and, consequently, on the strength of baryonic feedback.

In the right, we instead normalise the uncertainty by the optical depth profile corresponding to each gas distribution. In these \textit{relative} terms, the miscentering uncertainty is nearly independent of the underlying density profile. While some reduction in the differences is expected because the normalisation accounts for variations in signal amplitude, the near-complete cancellation of the dependence is less straightforward to explain. Whether this cancellation has a physical origin or is coincidental remains unclear and warrants further investigation.


\bibliographystyle{mnras}
\bibliography{main}
\bsp	
\label{lastpage}
\end{document}